\documentclass[pdflatex,sn-mathphys-num]{sn-jnl}

\usepackage{graphicx}%
\usepackage{multirow}%
\usepackage{amsmath,amssymb,amsfonts}%
\usepackage{amsthm}%
\usepackage{mathrsfs}%
\usepackage[title]{appendix}%
\usepackage{xcolor}%
\usepackage{textcomp}%
\usepackage{manyfoot}%
\usepackage{booktabs}%
\usepackage{algorithm}%
\usepackage{algorithmicx}%
\usepackage{algpseudocode}%
\usepackage{listings}%
\usepackage{layouts} 

\usepackage{subcaption}
\usepackage{siunitx}
\theoremstyle{thmstyleone}%
\theoremstyle{thmstyletwo}%

\theoremstyle{thmstylethree}%

\begin{document}

\title[Auditory and Visual Virtualization in 6DoF XR]{Auditory and Visual Virtualization in 6DoF XR: Effects on Plausibility and Dynamic Auditory Distance Perception}


\author*[1]{\fnm{Johannes M.} \sur{Arend}}\email{johannes.arend@aalto.fi}

\author[2]{\fnm{Christian} \sur{Scheer}}\email{christian.scheer@tu-berlin.de}

\author[3,2]{\fnm{Melissa} \sur{Ramírez}}\email{melissa.ramirez@th-koeln.de}

\author[2]{\fnm{Stefan} \sur{Weinzierl}}\email{stefan.weinzierl@tu-berlin.de}

\author[2]{\fnm{Fabian} \sur{Brinkmann}}\email{fabian.brinkmann@tu-berlin.de}

\affil*[1]{\orgdiv{Acoustics Lab, Department of Information and Communications Engineering}, \orgname{Aalto University}, \orgaddress{\street{Otakaari 5}, \city{Espoo}, \postcode{02150}, \country{Finland}}}

\affil[2]{\orgdiv{Audio Communication Group}, \orgname{Technische Universität Berlin}, \orgaddress{\street{Einsteinufer 17c}, \city{Berlin}, \postcode{10587}, \country{Germany}}}

\affil[3]{\orgdiv{Institute of Computer and Communication Technology}, \orgname{TH Köln - University of Applied Sciences}, \orgaddress{\street{Betzdorfer Str. 2}, \city{Cologne}, \postcode{50679}, \country{Germany}}}

\abstract{Extended reality (XR) enables controlled yet naturalistic multisensory experiments, but its validity depends on whether virtualized cues are perceptually plausible and lead to behavior comparable to the real world. To the best of our knowledge, this is the first study to perceptually evaluate a six-degree-of-freedom (6DoF) XR system that combines real-time room acoustic simulation, binaural rendering, and an audio-visually matched virtual replica of a large laboratory room in terms of both plausibility and behavior. In two experiments with 45 participants, we investigated how auditory and visual virtualization affect plausibility and dynamic auditory distance perception (ADP). A plausibility experiment compared real and virtual sound sources in audio-augmented reality (AAR) and virtual reality (VR), while a locomotion-based ADP experiment assessed distance-judgment accuracy, signed error, and task effort across four virtualization conditions. The system produced plausible binaural rendering in both AAR and VR, with higher plausibility in VR, indicating that perceptual requirements for XR audio depend on the visual reference frame. In the ADP task, any form of virtualization reduced accuracy and increased effort, although performance remained high overall. Virtual audio produced a bias toward nearer perceived source distances, whereas visual-only virtualization reduced accuracy without a significant directional bias. These effects did not add up in full VR. Overall, the findings demonstrate the strong potential of current 6DoF XR systems for interactive perceptual research, while revealing that task-specific behavioral validation may be necessary.}

\keywords{Extended Reality, Virtual Acoustics, Binaural Rendering, Multisensory Perception, Plausibility, Auditory Distance Perception}


\maketitle

\section{Introduction}\label{sec:Introduction}

Extended reality (XR) technologies, including augmented, mixed and virtual reality (AR, MR and VR), have had a significant impact on the design and conduct of perceptual multisensory studies and cognitive experiments~\cite{parsons_virtual_2015,apochi_virtual_2024,bello_extended_2025}. XR enables researchers to create a wide range of highly controlled yet naturalistic experimental environments that would be almost impossible to produce in traditional laboratory settings. This is particularly relevant to the fields of hearing science and psychoacoustics~\cite{picinali_vrar_2023,Ramirez2024_TIH,ramirez_speech--noise_2024,ramirez_screening_2026}, where the trade-off between experimental control and ecological validity has been recognized as a major limitation of traditional laboratory research~\cite{Keidser2020,beechey_ecological_2022}. Many auditory studies rely on simplified, unimodal, and static listening conditions that do not adequately reflect the complex, multisensory, and dynamic nature of real-world hearing~\cite{picinali_vrar_2023,Keidser2020,ramirez_screening_2026}. In contrast, immersive XR systems offer the ability to simulate everyday, interactive and multisensory environments---even for moving listeners and sources---while maintaining control over sensory cues. This elicits a sense of presence, leading participants to engage with the environment as if it were real and exhibit more close-to-natural perceptual and behavioral responses~\cite{Slater1997,Slater2009a,skarbez_immersion_2021}. 

Recent advancements in XR technology, particularly in room acoustic simulation and binaural rendering---the core technologies enabling spatialized audio in XR environments~\cite{Roginska2018,Vorlander2020,gupta_augmentedmixed_2022}---have significantly enhanced the realism and immersion of audiovisual simulations. These developments present promising opportunities for applying XR technology to auditory and hearing research. However, before state-of-the-art systems can be reliably used in complex multisensory experiments, their perceptual quality must be evaluated. Moreover, it is necessary to understand how auditory and visual virtualization affect behavior, and whether outcomes in XR are comparable to those in the real world. In particular, systematic perceptual evaluations of six-degree-of-freedom (6DoF) systems, in which users can move freely within a virtual space while the audio and visual renderings update continuously in real time, are lacking. Furthermore, the perceptual implications of auditory and visual virtualization in such 6DoF systems on perceptual quality measures, such as plausibility, and fundamental perceptual processes, such as dynamic auditory distance perception (ADP), remain poorly understood and have yet to be investigated.

\subsection{Plausibility Assessment}
One effective, rigorous, and popular method of evaluating the perceptual quality of virtual acoustic environments and binaural rendering is to assess the \textit{plausibility} of the environment using the experimental paradigm proposed by Lindau and Weinzierl~\cite{Lindau2012}. Effectively, such a plausibility experiment tests whether the XR audio rendering system can evoke auditory illusions, i.e., the impression that virtual sound sources rendered through headphones are actually real. In the experiment, either a real or a virtual sound source is presented, and participants decide whether the source is real or virtual using a Yes/No paradigm, thereby implicitly assessing whether the perceived sound is ``in agreement with the listener's expectation towards a corresponding real event''~\cite[p.~804]{Lindau2012}. This definition of plausibility in XR audio aligns with Slater's definition of \textit{plausibility illusion} in VR, which refers to the illusion that ``what is apparently happening is really happening''~\cite[p.~3553]{Slater2009a}. The results of an XR audio plausibility test are usually analyzed according to signal detection theory (SDT), yielding the sensitivity parameter $d'$, the corresponding two-alternative forced-choice detection rate (number of correct answers), and the response bias (e.g., criterion $c$ or likelihood ratio $\beta$) per participant~\cite{hautus_detection_2022}. These results provide a measure of the plausibility, and consequently, the perceptual quality of the assessed XR audio rendering, and are also assumed to offer insight into the system's general suitability for perceptual research.

Following the strict paradigm proposed by Lindau and Weinzierl~\cite{Lindau2012}, various studies have assessed the plausibility of binaural audio-augmented reality (AAR) rendering based on acoustic measurements. In the experiment introducing the paradigm, Lindau and Weinzierl~\cite{Lindau2012} yielded a very high degree of plausibility for binaural rendering with 1DoF, with participants mostly guessing whether a sound was real or simulated. The authors proposed that a binaural AAR system can be considered plausible if the mean $d'$ across subjects is significantly smaller than a threshold of $d' = 0.1777$, as determined by a minimum effect test of the null hypothesis. Although this threshold was commonly used in later studies, it represents a rather conservative measure of plausibility, reflecting only a 5\% deviation from chance-level performance in a two-alternative forced-choice (2AFC) task. Therefore, other studies~\cite{Brinkmann2019a} have proposed more lenient thresholds, such as $d' = 0.9539$, which corresponds to 75\% correct responses in a 2AFC task and is often considered a practical definition of the just-noticeable difference (JND). Correcting the detection rate for the guessing probability shows that, at the JND level, participants are giving the correct answer in 50\% of the cases. This criterion may hence better reflect the perceptual boundary at which AAR rendering begins to feel convincingly real.

Following the described paradigm, Pike et al.~\cite{Pike2014} assessed the plausibility of a 1DoF binaural rendering system in a smaller room and observed small yet meaningful sensory differences, indicating a slightly lower degree of plausibility (higher detection rates) of their system compared to that of Lindau and Weinzierl, even though participants still found it difficult to identify the rendering. Using 6DoF parametric binaural rendering based on a single monaural room impulse response measurement, Arend et al.~\cite{Arend2021c} conducted a pilot study on the plausibility of their system, in which participants could freely move within a listening area, and found that the rendering could be considered plausible for one out of four expert participants. Neidhardt and Zerlik~\cite{Neidhardt2021} further explored the plausibility of position-dynamic 2DoF binaural rendering, whereby listeners' translational movements were restricted to walking along a line. The authors concluded that, when applying Lindau and Weinzierl's definition, their evaluated system did not yield a plausible rendering for all participants. Lübeck et al.~\cite{lubeck_perceptual_2024} systematically investigated the influence of various rendering parameters on binaural rendering in anechoic conditions, yielding plausible results with their baseline system. Notably, the visibility of the real sound sources (loudspeakers) and the ability of seated participants to perform translational head movements had only a minimal impact on plausibility. Himmelein et al.~\cite{Himmelein2025} extended these investigations by comparing the plausibility of 3DoF binaural rendering in a reverberant office environment based on dummy head measurements, microphone array measurements, and---as a first addition to measurement-based approaches---parametric simulation. The authors obtained plausible results for their binaural rendering and found no significant differences in overall plausibility between the three conditions, suggesting that the specific choice of rendering method has only a minor impact. Last, Meyer-Kahlen et al.~\cite{MeyerKahlen2024} compared different experimental paradigms for assessing the capability of measurement-based static binaural rendering (i.e., 0DoF) to evoke auditory illusions, including the Lindau and Weinzierl plausibility Yes/No task. Even their most technically accurate system exceeded the threshold proposed by Lindau and Weinzierl, suggesting that the rendering would not be deemed plausible. Nevertheless, the authors demonstrated that the plausibility paradigm effectively reveals quality differences between various rendering conditions, highlighting its effectiveness in assessing the overall quality of XR audio rendering.

Although it is a valuable method for evaluating the quality of binaural rendering, the studies employing this paradigm to date have focused primarily on measurement-based binaural rendering systems for AAR. Furthermore, many of these studies have restricted listener movement, ultimately limiting ecological validity and reducing resemblance to everyday listening experiences. Yet, the plausibility of current 6DoF real-time room acoustic simulation and binaural rendering engines has not been evaluated. This leaves the quality of state-of-the-art XR audio systems, which are typically based on real-time room acoustic simulation in which sound propagation is dynamically computed, unclear. Furthermore, existing plausibility studies have only focused on AAR settings, in which the visual scene remains real. Thus, the impact of combined auditory and visual virtualization---as in fully immersive VR---on the plausibility of XR audio rendering has yet to be systematically investigated.

\subsection{Effect of Virtualization on Auditory Distance Perception}

While assessing plausibility provides valuable insights into the perceptual quality of XR audio rendering, it remains unclear whether a plausible environment necessarily guarantees behavior equivalent to that observed in the real world---a critical factor when using XR environments for behavioral research on multisensory perception. In particular, little is known about how auditory and visual virtualization, even if plausible, affects fundamental perceptual processes. One such process is auditory distance perception (ADP), which plays a crucial role in real-world spatial hearing and depends on the complex interplay of acoustic, visual, and sensory-motor cues~\cite{Zahorik2005,Kolarik2016}. To date, empirical comparisons of ADP performance in real versus virtual environments are limited, especially regarding \textit{dynamic} ADP in immersive 6DoF scenarios. Additionally, while several recent studies have investigated the effect of virtualization on ADP under static or at most 3DoF conditions, the virtual environments were not evaluated for perceptual plausibility beforehand. This limits the generalizability of their findings and leaves the baseline fidelity of the XR audio system unclear, making it uncertain whether the observed effects stem from virtualization itself or suboptimal rendering quality.

In unimodal (audio-only) conditions, ADP of far-field sound sources primarily relies on the combination of the well-established acoustic cues sound intensity, direct-to-reverberant energy ratio (DRR), and spectral content~\cite{Zahorik2005,Kolarik2016}. Sound intensity typically decreases with distance according to the inverse-square law, even though in reverberant spaces, the attenuation rate is reduced due to the (almost) distance-independent diffuse sound field that is generated by reflections. DRR, representing the balance between direct and reflected sound, is another important cue in reverberant environments, decreasing as source distance increases. Spectral cues emerge for far-away sources (more than 15 meters away), as air absorption causes high-frequency attenuation. Especially at shorter distances, dynamic listening conditions introduce additional cues, such as acoustic tau, which is the rate of change in sound level as the listener moves, and motion parallax, which is the change in source direction as the listener moves. In multimodal (audiovisual) conditions, ADP is significantly influenced by visual information. This typically causes the perceived auditory source to be pulled toward a visible target, an effect known as visual capture or the ventriloquism effect~\cite{Alais2004,zahorik_asymmetric_2022}. The effect occurs because visual estimates of distance are generally more reliable than auditory ones, likely resulting in visual dominance in multimodal perception~\cite{Kolarik2016}.

Auditory or visual virtualization can alter key distance cues, thereby affecting ADP in XR environments. For instance, inaccuracies in room acoustic simulation may result in DRR values that differ from those of the real room, causing a shift in perceived auditory distance. Similarly, wearing a head-mounted display (HMD) typically causes visual egocentric distances to be underestimated, compressing them to approximately 74–82\% of their actual modeled size in VR~\cite{kelly_distance_2023, kurpinar_strengths_2025}. This visual compression can interact with visual capture mechanisms, potentially distorting ADP in VR environments.

Research on the potential effects of virtualization on ADP is still very scarce, and to date, very few studies have directly compared ADP between real and virtual rooms. Blau et al.~\cite{Blau2021} compared speech stimuli played over loudspeakers in a real room with their virtual reproduction using 1DoF binaural rendering based on measured and offline simulated binaural room impulse responses (BRIRs). In an AAR setting, listeners compared the (visible) real to the virtual sources in terms of their consistency with regard to various spatial attributes, including ADP, on a nine-point quality scale. The authors found that renderings based on measured BRIRs were rated as strongly consistent with the real room for most attributes, whereas renderings based on simulated BRIRs were considered convincing but received lower ratings. In a similar manner, Stärz et al.~\cite{starz_comparison_2025} compared speech stimuli played over real and virtual loudspeakers in a VR setting in which participants saw a virtual representation of the room they were in. The authors employed 1DoF binaural rendering based on measured or offline simulated individual or non-individual BRIRs for virtualizing the loudspeaker. Participants compared the real and virtual loudspeakers based on seven room acoustic qualities, including source distance, using a 100-point scale. The results showed that, for most of the examined rendering options, the perceived source distance was comparable to that of the real loudspeaker, indicating good overall agreement for ADP between the real and virtual environments. Stodt et al.~\cite{stodt_comparing_2024} compared the detection of changes in sound source distance in real and virtual reverberant environments using an oddball paradigm along with electroencephalography (EEG) measures. For their AAR setup, the researchers used static binaural rendering (i.e., 0DoF), based on measured dummy head BRIRs, to virtualize the visible loudspeakers in the room. The results showed that, although neurocognitive processing patterns were comparable in real and virtual settings, detecting distance changes was reduced with the virtualized loudspeakers. This was particularly evident when the dominant loudness cue was altered to match across distances, underscoring the critical role of accurate loudness representation for good task performance in XR audio. Lastly, Lombera et al.~\cite{lombera_comparison_2025} compared ADP of real and virtual sound sources, and examined the effect of acoustic congruence between the recording and reproduction rooms. Participants were blindfolded, and the AAR reproduction was performed with static binaural rendering (i.e., 0DoF) based on measured individualized or non-individualized BRIRs, which prevented visual and dynamic auditory cues. Their results showed that both individualized and non-individualized BRIRs yielded ADP comparable to that of real sound sources in a congruent listening environment. However, incongruence in the form of a significantly lower reverberation time in the reproduction room than in the recording room compressed perceived distances for at least a subset of participants, indicating room divergence effects~\cite{Werner2016,Amengual2021} that must be considered for XR audio rendering.

Complementing these studies on auditory virtualization, other recent work has examined how virtualizing or manipulating the visual context affects ADP while keeping the auditory scene physically real. Roßkopf et al.~\cite{roskopf_impact_2025} examined multisensory effects on ADP by altering the visual scene while keeping the auditory scene unchanged. In their study, participants standing still and wearing an HMD estimated the distance of real loudspeakers while viewing different virtual room models. These models were either congruent or incongruent with the real room in terms of size, and provided either congruent or very limited spatial information by presenting only a loudspeaker in an otherwise dark environment. Participants indicated perceived distances through a placement task, by walking, or verbally. The study found that audiovisual room incongruence negatively affected ADP, leading to a general overestimation of distances in VR that was notably higher with the placement task in the dark environment. These results indicate clear effects of the visual scene on auditory perception in VR. Cerviño et al.~\cite{cervino_influence_2026} further investigated whether virtual visual context can modulate ADP of real sound sources in a mixed-reality setting. Across four experiments, seated participants estimated the distance of real loudspeakers while visual information was manipulated either by contrasting real visual information with complete darkness, or by presenting virtual room models through an HMD, including rooms of different perceived sizes. The results showed that real visual information improved ADP relative to darkness, whereas virtual visual environments did not reliably affect auditory distance estimates, even when virtual room size was manipulated or perceived as substantially larger. This suggests that virtual visual context does not necessarily reproduce the audiovisual influence on ADP observed in real physical environments. 

Taken together, these studies have advanced our understanding of ADP in virtualized settings, but their scope has been limited. None have examined dynamic, locomotion-based ADP, where listener movement during stimulus presentation provides continuous distance updates that more accurately reflect real-world listening and improve accuracy through motion-related cues~\cite{ashmead_contribution_1995}, similar to how visual distance estimates in VR improve through motion cues~\cite{waller_correcting_2008,siegel_walking_2017}. Moreover, no prior work has systematically investigated auditory and visual virtualization in a unified, controlled paradigm, nor has any study assessed dynamic ADP in plausible, state-of-the-art, 6DoF XR environments with real-time room acoustic simulation. Addressing these gaps is essential to understanding whether recent XR systems can accurately reproduce everyday multisensory spatial hearing and can be effectively used in behavioral and perceptual research. 

\subsection{Present Study and Contributions}

Based on the research gaps identified above, this study addresses four key open research questions. RQ1: Can a state-of-the-art 6DoF XR audio system based on real-time room acoustic simulation provide plausible binaural rendering? RQ2: Does visual virtualization---transitioning from AAR to VR---affect plausibility and thereby indicate different perceptual requirements for AAR and VR audio rendering? RQ3: How do auditory and visual virtualization individually affect the fundamental perceptual processes of dynamic ADP in a locomotion-based task? RQ4: Do the effects of auditory and visual virtualization combine additively in the fully virtual condition, or is performance primarily determined by one modality?

To answer these questions, we conducted a two-part listening experiment in a mixed-reality laboratory using a custom 6DoF XR system combining liveRAZR~\cite{Schutte2021_Dissertation,de_haas_real-time_2025} real-time room acoustic simulation and binaural rendering with Unreal Engine-based visual rendering. First, participants completed a plausibility test in AAR (i.e., virtual audio and real vision) and VR (i.e., virtual audio and virtual vision) contexts. Second, they performed a dynamic, locomotion-based ADP task under four virtualization conditions---real audio~/~real vision, real audio~/~virtual vision, virtual audio~/~real vision, and virtual audio~/~virtual vision---allowing us to systematically examine modality-specific and combined virtualization effects. The main contributions of this paper are as follows:
\begin{itemize}
    \item We present the technical implementation and calibration of a real-time 6DoF XR system for perceptual experiments, combining real-time room acoustic simulation and binaural rendering with a visually detailed virtual replica of a large laboratory room.
    \item We provide a plausibility evaluation of the 6DoF binaural rendering in both AAR and VR, demonstrating how the visual reference frame influences the perceived plausibility of XR audio rendering.
    \item We evaluate dynamic ADP in a locomotion-based task under independent auditory and visual virtualization, revealing distinct modality-specific effects on distance-identification accuracy, signed response error, and self-reported effort.
    \item We show that plausibility does not necessarily imply task-specific behavioral equivalence, thereby motivating combined perceptual quality and behavioral validation procedures for XR systems used in multisensory perception research.
\end{itemize}

\section{General Method}
The two experiments reported in this paper---the plausibility assessment and the dynamic ADP study---were both conducted within the same overarching experimental protocol. All participants completed both experiments across two sessions on separate days, with tasks interleaved and counterbalanced to control for potential order effects. Since the two experiments share elements such as participants, the laboratory setup, and the XR system, these aspects are described first in the following sections. The experiment-specific stimuli, procedure, results, and discussion are presented separately for clarity.

\subsection{Participants}
A total of 45 participants (ages 21–42 years, M = 27.71 years, Mdn = 26 years, SD = 4.97) voluntarily participated in the experiment. Two participants reported having mild hearing problems in their right ear, the rest had self-reported normal hearing. Most were master’s students in Audio Communication and Technology at Technische Universität Berlin, some were students from other fields, and some were members of our laboratory. Twenty-nine of the participants had previously performed listening experiments. On average, the participants spent 3.8 hours per day listening to, playing, or working with audio. All participants were naive as to the purpose of the study.

\subsection{Setup}

\begin{figure*}[ht]
    \centering
    \includegraphics[width=1\textwidth]{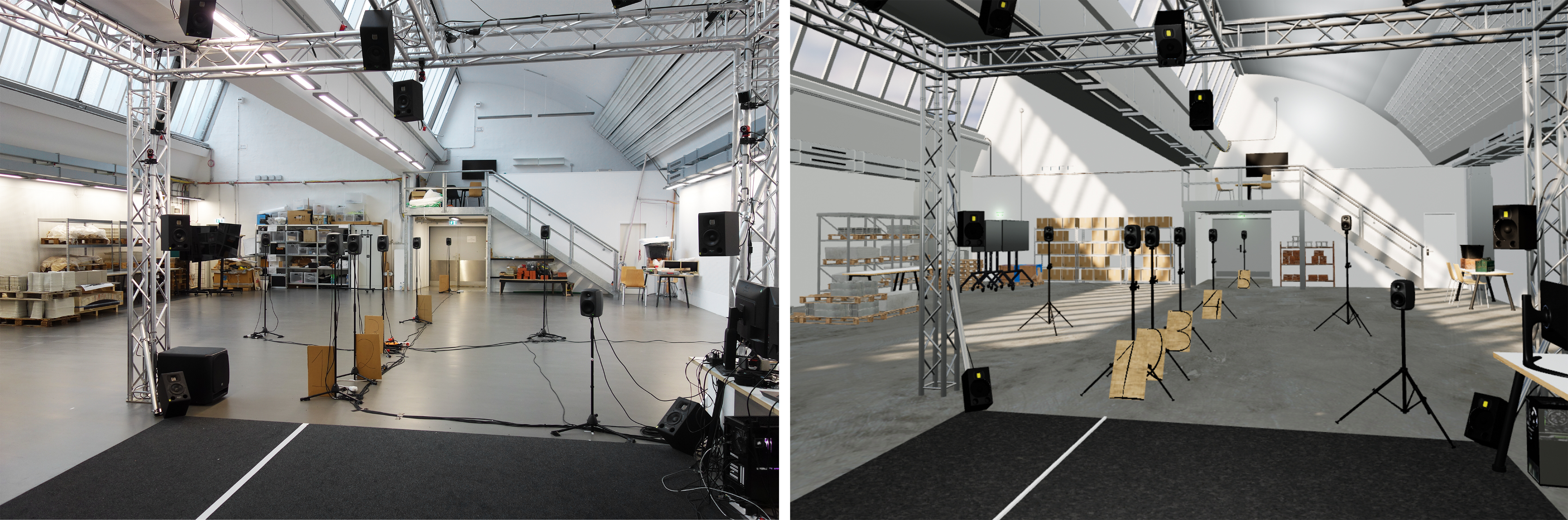}
    \caption{Photo of the Mixed Reality Lab showing the view from the listening area into the room with the eight Genelec loudspeakers (left). Screenshot showing the real-time rendered HMD view of the virtual visual representation of the laboratory room and experimental setup (right).}
    \label{fig:MRL_Real_Unreal}
\end{figure*}

\subsubsection{Experimental Environment}
We conducted the experiment in the Mixed Reality Lab of Technische Universität Berlin and Universität der Künste Berlin. The space is relatively large, with a main usable area of approximately \SI{287}{\square\meter}, a gallery, ceiling windows, and vaulted ceilings that vary in height from about \SI{3.60}{\metre} to \SI{6.60}{\metre}. The room is acoustically treated with perforated plasterboard panel absorbers covering the side and back walls. It has a mid-frequency reverberation time of \SI{1.21}{\second} and a background noise level of about \SI{42.3}{\decibel (A)}. Figure~\ref{fig:MRL_Real_Unreal} (left) shows a photo of the laboratory room, and the supplementary material~\cite{supp_material} contains additional pictures (Figs.~S1 to S6).

\begin{figure}[b]
    \centering
    \includegraphics[width=.5\textwidth]{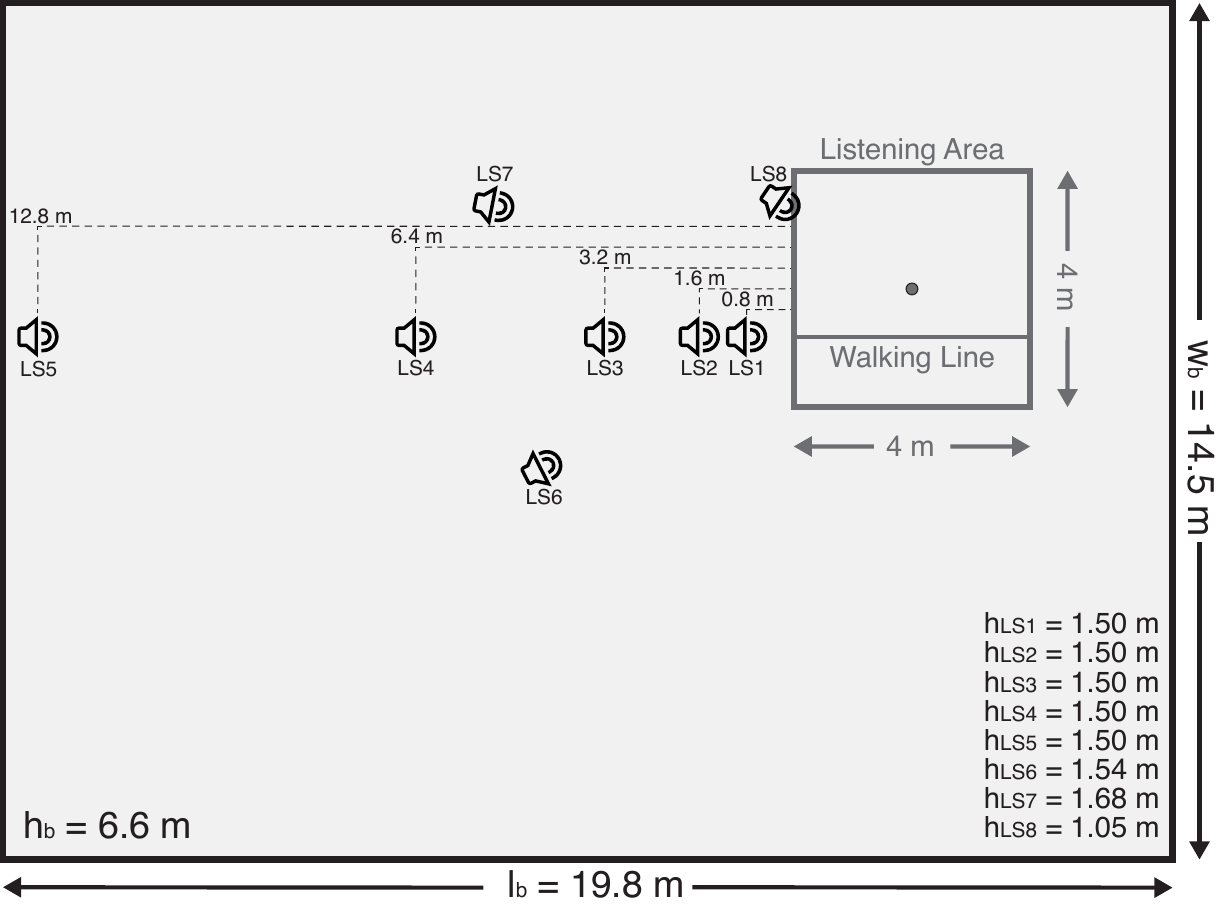}
    \vspace{1mm}
    \caption{Floor plan of the experimental environment.}
    \label{fig:floor_plan}
\end{figure}

The laboratory has a square listening area of 4 $\times$ \SI{4}{\meter}, indicated by a black carpet. The area is surrounded by an aluminum truss system on which tracking cameras and loudspeakers of an Ambisonics array are mounted. The latter was not used in this study. Instead, we used a total of eight Genelec 8020D loudspeakers, placed in front of the listening area. Five of the loudspeakers were arranged in a straight line with logarithmic spacing at distances between 0.8 to \SI{12.8}{\meter} from the edge of the listening area (i.e., 0.8, 1.6, 3.2, 6.4, and \SI{12.8}{\meter}) at a height of \SI{1.50}{\meter} measured at the loudspeaker's acoustical axis. These loudspeakers, denoted LS1 to LS5, were used for both experiments but were positioned with logarithmic spacing to facilitate the distance perception experiment; logarithmic spacing produces more uniform perceptual steps across the tested range, consistent with the logarithmic scaling of auditory distance perception and its acoustic cues~\cite{Zahorik2005,zahorik_asymmetric_2022}. To increase variability in the sources for the plausibility study, we placed three additional loudspeakers (LS6 to LS8) to the left and right of the center of the listening area at heights of 1.54, 1.68, and \SI{1.05}{\meter}, respectively. LS8 was positioned particularly low and close to the listening area so that participants could experience sources at different heights very close to them. LS1 to LS5 pointed toward the walking line, which was a white line on the carpet indicating the path participants had to walk along for the distance study, while LS6 to LS8 pointed toward the center of the listening area. Figure~\ref{fig:floor_plan} illustrates the described setup, showing the dimensions, the listening area (including the center point and walking line), and the position and height of the loudspeakers.

\subsubsection{XR System}

\begin{figure*}[ht]
    \centering
    \includegraphics[width=1\textwidth]{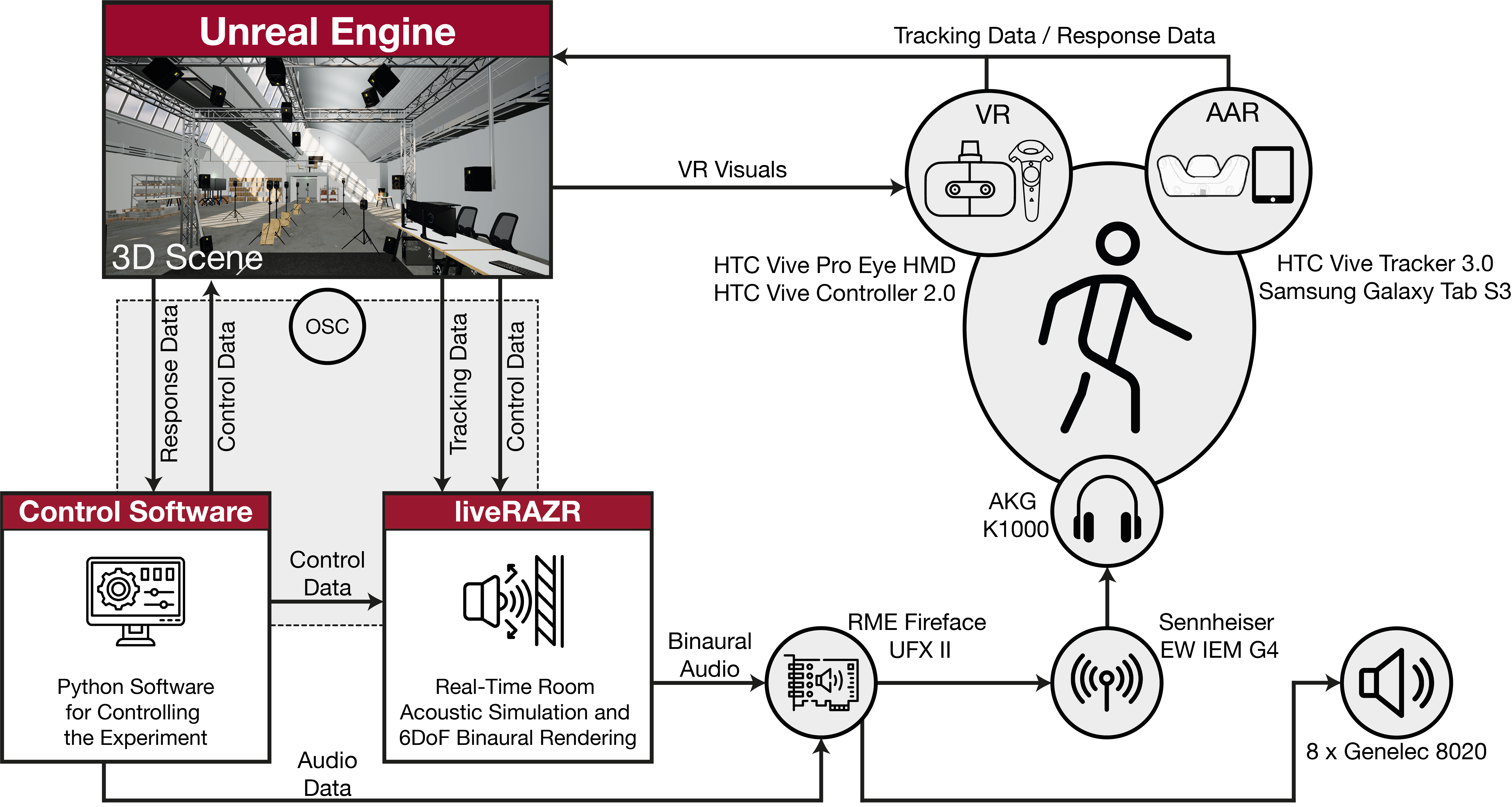}   
    \caption{Overview of the XR system.}
    \label{fig:XRsystem}
\end{figure*}

The purpose-built XR system is a modular combination of various hardware and software components, as illustrated in Fig.~\ref{fig:XRsystem}. In terms of software, we used the game and XR development platform Unreal Engine 5.1.1 for visual rendering, as well as for gathering and distributing tracking, control, and response data using OpenXR and Open Sound Control (OSC). In particular, Unreal received listener tracking data from the XR hardware, rendered the visual scene accordingly, and transmitted listener and sound source position/orientation data via OSC to liveRAZR 0.6.0, which we employed for real-time room acoustic simulation and 6DoF binaural rendering~\cite{Schutte2021_Dissertation,Sommerfeldt2024,de_haas_real-time_2025}. Last, we implemented Python software for controlling the experiment. This software exchanged control data with Unreal Engine and liveRAZR via OSC to activate virtual sound sources and specify audio content to be played. It also sent audio signals to the loudspeakers and received participants’ responses.

Regarding XR hardware, we used an HTC Vive Pro Eye HMD (without the headphones) together with an HTC Vive Controller 2.0 to provide responses in VR contexts, or a HTC Vive Tracker 3.0 (attached to the employed headphones; see below) combined with a Samsung Galaxy Tab S3 for user input in AAR contexts. Four SteamVR Base Stations 2.0 mounted on the aluminum truss system were used for tracking the listening area, running at a tracking rate of \SI{120}{\Hz}. The SteamVR 2.3.5 software handled the Vive and SteamVR hardware. In terms of audio hardware, we employed AKG K1000 extra-aural headphones to present the binaural signals. These headphones are often used in XR audio research due to their acoustically transparent design, which only minimally impairs the real sound field~\cite{Porschmann2019d}. The headphones were connected to a Sennheiser EW IEM G4 wireless audio transmitter, enabling users to move freely within the listening area. The monitoring system and the eight Genelec loudspeakers were fed by an RME Fireface UFX II audio interface, which operated at a sampling rate of \SI{48}{\kHz} and a buffer size of 512 samples. The entire XR system ran on a Windows 10 Pro PC with 64 GB of memory, a 16-core AMD Ryzen 9 5950X CPU, and an NVIDIA GeForce RTX 3070 GPU.

Depending on the experimental condition, participants' response data was collected in different ways. In VR scenarios (or, more generally, in any condition in which participants wore the HMD), an interactive menu appeared in the 3D scene. This allowed participants to submit their responses directly with the Vive controller inside the VR space. In AAR scenarios (or, more generally, in any condition where participants did not wear the HMD and thus had real vision), participants submitted their responses using the Samsung tablet displaying a simple graphical user interface (GUI) created with TouchOSC, which sent the responses via OSC.

\subsubsection{Visual Rendering}
The 3D model of the Mixed Reality Lab was designed in Blender 4.0 and imported as an FBX file into Unreal Engine 5.1.1. Additional objects from the Quixel library were added to the model to create a scene resembling the real room and providing sufficient visual depth and distance reference cues (see Figs.~S7 to S11 in the supplementary material~\cite{supp_material} for screenshots of the Unreal model). The VR visuals were rendered in real time through the OpenXR interface for the HTC Vive Pro Eye HMD at its native resolution of 1440 $\times$ 1600 pixels per eye with a 110-degree field of view. Unreal Engine's frame rate was uncapped to ensure smooth real-time performance, though the HMD displayed at its native \SI{90}{\Hz} refresh rate. The rendering employed forward shading with 4 $\times$ multisample anti-aliasing (MSAA), instanced stereo rendering for both eyes in a single GPU pass, and all lighting was static and baked into lightmaps to reduce runtime costs. This configuration helped maintain high frame rates while delivering stable images, which is crucial for VR experiences. Figure~\ref{fig:MRL_Real_Unreal} (right) shows a screenshot of the real-time rendered HMD view (see Figs.~S12 and S13 in the supplementary material~\cite{supp_material} for further screenshots of the HMD view).

\subsubsection{Audio Rendering}
We used liveRAZR 0.6.0~\cite{Schutte2021_Dissertation,Sommerfeldt2024,de_haas_real-time_2025} for real-time room acoustic simulation of the Mixed Reality Lab and 6DoF binaural rendering. This state-of-the-art C++ software is optimized for efficient simulation and rendering while maintaining physical accuracy, enabling artifact-free and high-quality rendering of moving sources and receivers, which makes it perfectly suited for 6DoF XR systems.

The room acoustic simulation in liveRAZR 0.6.0 is based on a rectangular (shoebox) room with an image source model (ISM) for direct sound and early reflections, and a feedback delay network (FDN) for late reverberation. Sound source directivity is integrated as part of the ISM, where the direct sound and each image source are filtered with the respective directivity filter. For binaural rendering, the room acoustic simulation is spatialized using a set of head-related transfer functions (HRTFs). Both the directivity filters and the HRTFs are loaded from SOFA files~\cite{Majdak2022} and are interpolated for the required directions using vector base amplitude panning (VBAP)~\cite{Pulkki1997}.

The ISM was set to a maximum order of two, with edge diffraction filters and surface scattering enabled, as well as fractional delay filters applied to the propagation delay. The FDN consisted of 12 channels with delays scaled according to the room dimensions, as well as a randomly generated orthogonal feedback matrix, both of which are default settings. As source directivity filters, we used the Genelec 8020D directivity measurements from the BRAS database~\cite{Brinkmann2021}. We subsampled the spatially ultra-dense measurement set to 974 directions (Lebedev grid of order 26) and converted the impulse responses to minimum-phase FIR filters with a fairly large tap size of 2048 samples to accurately represent the loudspeaker's low-frequency roll-off. For the HRTFs, we used the 2702-direction Neumann KU100 dataset by Bernschütz~\cite{Bernschutz2013a}, and subsampled it to 900 directions (Fliege-Maier grid of order 29). We time-aligned the subsampled dataset by extracting the time-of-arrivals (TOAs) using onset detection; a preprocessing step necessary to prevent comb-filter artifacts when interpolating HRTFs~\cite{Arend2021a,Arend2023b}. The HRTFs were saved as a SOFA file with a filter length of 128 taps and the TOAs stored in the data structure so that liveRAZR can correctly reinsert the TOAs after interpolation. Additionally, to minimize the influence of the headphones used, we applied a headphone compensation filter for the combination of the Neumann KU100 HRTFs and the AKG K1000 headphones. This filter was based on twelve measurements (putting the headphones on and off the dummy head) to account for repositioning variability, and it was designed using regularized inversion of the complex mean of the headphone transfer functions~\cite{Erbes2017}. We applied the filter offline to all input signals to avoid additional real-time computations.

\subsubsection{Room Acoustic Matching}
As mentioned above, liveRAZR 0.6.0 uses shoebox-shaped rooms with six surfaces rather than precise geometries, making the simulation much more computationally efficient. Various studies have demonstrated that shoebox-based room acoustic simulations yield high-quality perceptual results and that the auditory accuracy achieved through this approximation is sufficient for many applications~\cite{Brinkmann2019a,Blau2021,starz_comparison_2025, Himmelein2025}. Thus, using a shoebox-based simulation aligned with our rationale of using a lightweight yet high-quality solution for the 6DoF XR system. In the present case, approximating the Mixed Reality Lab's room geometry with a shoebox primarily simplified the ceiling structure because the rest of the room is nearly rectangular. Setting the model's height equal to the maximum ceiling height resulted in a shoebox-shaped model measuring 19.8 $\times$ 14.5 $\times$ \SI{6.6}{\meter} (L $\times$ W $\times$ H; see Fig.~\ref{fig:floor_plan}).

\begin{figure*}[t]
    \centering
    \begin{minipage}[h]{0.315\textwidth}
        \centering
        \includegraphics[width=\textwidth]{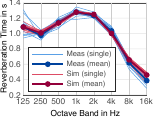}
    \end{minipage}
    \hfill
    \begin{minipage}[h]{0.325\textwidth}
        \centering
        \includegraphics[width=\textwidth]{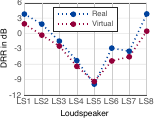}
    \end{minipage}    
    \hfill
    \begin{minipage}[h]{0.325\textwidth}
        \centering
        \includegraphics[width=\textwidth]{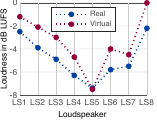}
    \end{minipage}  
    \caption{Frequency-dependent reverberation time of the real and simulated room, with thin lines showing single measurements or simulations and thick lines with markers showing the mean values (left). Broadband DRR (center) and normalized loudness (right) of the eight real and virtual loudspeakers at the center of the listening area.}
    \label{fig:T20_DRR_LUFS}
\end{figure*}

With the geometry of the shoebox room established, the next step was to match its acoustic characteristics to those of the real room. In line with recent XR audio developments, our matching approach was to adjust the simulation based on global room acoustic parameters rather than attempting to fine-tune individual sound sources. To match reverberation time, we initialized frequency dependent absorption coefficients for each wall using tabulated data (e.g., perforated plasterboard panel absorbers and concrete floor), measured and simulated 12 room impulse responses (2 loudspeaker $\times$ 6 microphone positions), computed the reverberation time in octave bands between \SI{125}{\Hz} and \SI{16}{kHz}, and adjusted the initial absorption coefficients iteratively. For the measurements, we used a Genelec 8331A loudspeaker and an NTI M2211 Class 1 microphone.

Throughout this process, we maintained the room's inhomogeneous absorption behavior, i.e., the concrete floor was barely absorbent, while the paneled walls were strongly absorbent, especially at higher frequencies. Figure~\ref{fig:T20_DRR_LUFS} (left) shows the frequency-dependent reverberation time for each of the twelve single measurements or simulations, as well as the respective means. The plot reveals a high degree of similarity between the average reverberation times of the real and virtual rooms, with only minor differences observed at higher frequencies in the \SI{16}{\kHz} band. These results suggest high perceptual congruence between the real and virtual rooms with respect to reverberation time.

When comparing acoustic parameters of the real and virtual rooms, we observed a slight mismatch in DRR, with the simulation tending to produce higher values. To estimate the necessary adjustments, we placed a Neumann KU81i dummy head at an ear-height of \SI{1.6}{\meter} in the center of the listening area (see Fig.~\ref{fig:floor_plan}) and measured BRIRs of all loudspeakers using the swept-sine method. The sweep was either (i) played through the real loudspeakers in the room or (ii) routed to liveRAZR and the virtual loudspeakers, in which case the resulting binaural signal was reproduced via the AKG K1000 headphones worn by the dummy head (see Fig.~S1 and Fig.~S2 in the supplementary material~\cite{supp_material}). Based on these measurements, we increased the overall reverberation level of the simulation by increasing the gain of the ISM (except direct sound) and FDN by \SI{4}{\decibel}, which resulted in a close match of DRR between real and virtual loudspeakers at the tested position (see Fig.~\ref{fig:T20_DRR_LUFS}, center). The only notable deviation occurred for the relatively close and low LS8, for which a difference of about \SI{3.4}{\decibel} remained. This is close to the reported average JND of 2.4 to \SI{3.8}{\decibel} for this DRR range~\cite{Larsen2008}, suggesting that differences in DRR would even be difficult to detect in a direct comparison between the real and virtual sound sources---a situation that never occurred in our experiments. Figure~S14 in the supplementary material~\cite{supp_material} provides further details, showing the frequency-dependent DRR of the real and virtual loudspeakers at the listening area center position. The figure reveals that the slightly higher DRR of the real loudspeakers is mostly driven by the octave bands below \SI{1}{\kHz} and the high-frequency band of \SI{16}{\kHz}.

In a last step, we adjusted the playback level of the real loudspeakers to an appropriate range for the experiment and matched the loudness of the real and virtual loudspeakers. The real loudspeakers produced levels of approximately \SIrange{56}{72}{dBLAeq}, depending on the loudspeaker and microphone position in the listening area, measured with pink noise and an NTI XL2 with an NTI M2211 Class 1 microphone at \SI{1.60}{\meter} ear-height. For loudness matching, we recorded pink noise reproduced by real and virtual loudspeakers with a Neumann KU81i dummy head at various positions across the listening area, computed the ITU-R BS.1770-4 loudness, and iteratively adjusted the headphone level until the overall loudness match was satisfactory. Figure~\ref{fig:T20_DRR_LUFS} (right) shows the normalized loudness of the real and virtual loudspeakers at the center of the listening area. The plot reveals good overall agreement, though the virtual loudspeakers have slightly higher loudness, with a maximum deviation of about 2~dB~LUFS for LS8. As with the DRR deviations, these differences may be audible in a direct comparison; however, participants could not make such comparisons, and the differences varied across the listening area. We therefore consider it unlikely that the small loudness differences affected the study.

\begin{figure*}[t]
    \centering
    \begin{minipage}[h]{0.49\textwidth}
        \centering
        \includegraphics[width=\textwidth]{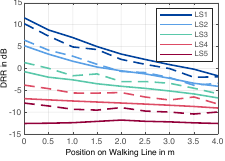}
    \end{minipage}
    \hfill
    \begin{minipage}[h]{0.49\textwidth}
        \centering
        \includegraphics[width=\textwidth]{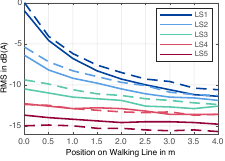}
    \end{minipage}    
    \caption{DRR (left) and normalized level (right) of the real (solid lines) and virtual (dashed lines) loudspeakers LS1 to LS5 along the walking line.}
    \label{fig:BOL_DRR_RMS_RvsV}
\end{figure*}

With the room matching complete, we evaluated how accurately the system reproduced the main dynamic ADP cues---DRR and level changes---along the walking line used in the ADP study. Similar to the procedure described above, we used the Neumann KU81i dummy head at \SI{1.60}{\meter} ear-height to measure DRR and A-weighted RMS levels of the first five real and virtual loudspeakers in \SI{0.50}{\meter} steps along the walking line. Figure~\ref{fig:BOL_DRR_RMS_RvsV} shows the results for DRR (left) and normalized level (right) whereby the \SI{0}{\meter} position on the walking line denotes the location in front of LS1 (at a distance of \SI{0.8}{\meter}; see Fig.~\ref{fig:floor_plan}). The results show that DRR and levels of real and virtual loudspeakers are concurrent and generally decrease as distance increases, with stronger effects for the closer loudspeakers LS1 and LS2 and almost no change for the far-away loudspeaker LS5. Regarding the DRR, Fig.~\ref{fig:BOL_DRR_RMS_RvsV} (left) reveals slight difference between the real and virtual loudspeakers, especially for LS5. However, the absolute differences in all measurements do not exceed the JND in the respective DRR range~\cite{Larsen2008}, indicating good overall perceptual similarity with adequate dynamic DRR cues. For a more detailed analysis, Fig.~S15 and S16 in the supplementary material~\cite{supp_material} show the absolute broadband DRR differences and the frequency-dependent DRR across the walking line, respectively. The first plot provides a good estimation of the absolute difference with respect to the JND for a given DRR range~\cite{Larsen2008}. The second plot shows that DRR differences are largely driven by higher frequencies in octave bands from \SI{4}{\kHz} upwards. Regarding the level shown in Fig.~\ref{fig:BOL_DRR_RMS_RvsV} (right), besides exhibiting a similar distance-dependent behavior, the virtual versions of LS1 to LS3 appear to be slightly higher in level than their real counterparts, while the opposite occurs for LS5. However, the absolute level differences mostly do not exceed \SI{1}{\decibel}, with a maximum difference of \SI{1.3}{\decibel}, which clearly shows that the simulation adequately reproduces dynamic level cues similar to those in the real room. 

\subsection{Experimental Design}

\begin{figure}[htb]
    \centering
    \includegraphics[width=.5\textwidth]{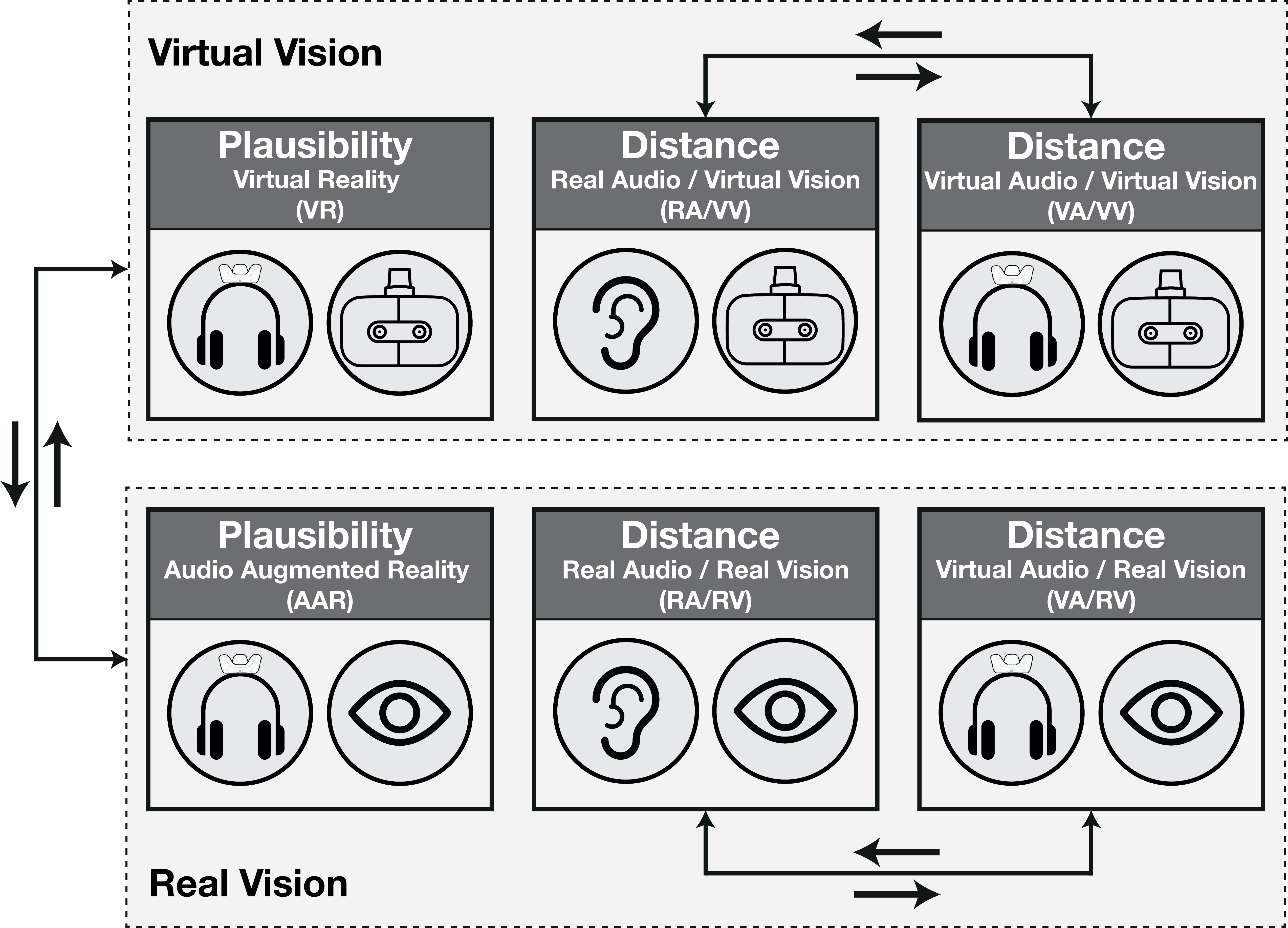}
    \caption{Illustration of the experimental design showing the different conditions for the plausibility and ADP experiments. The arrows indicate counterbalancing.}
    \label{fig:procedure}
\end{figure}

Participants performed the study over the course of two days, carrying out one plausibility experiment and two ADP experiments each day. Figure~\ref{fig:procedure} provides an overview of the experimental design. The study was counterbalanced across participants with regard to the visual conditions so that half of the participants started with the virtual vision block and the other half with the real vision block (upper and lower blocks in Fig.~\ref{fig:procedure}). On each day, participants performed the three experiments per block, starting with the plausibility experiment, followed by two distance perception experiments, which were also counterbalanced. Including instructions from the experimenter, procedural training sessions before each experiment, and sufficient breaks, the three experiments per day took about \SIrange{1}{1.25}{h} (also see the procedure descriptions in Sec.~\ref{sec:Exp1} and \ref{sec:Exp2} for more information).

\begin{figure*}[htb]
    \centering

    \begin{minipage}[t]{0.245\linewidth}
        \centering
        \includegraphics[width=\linewidth]{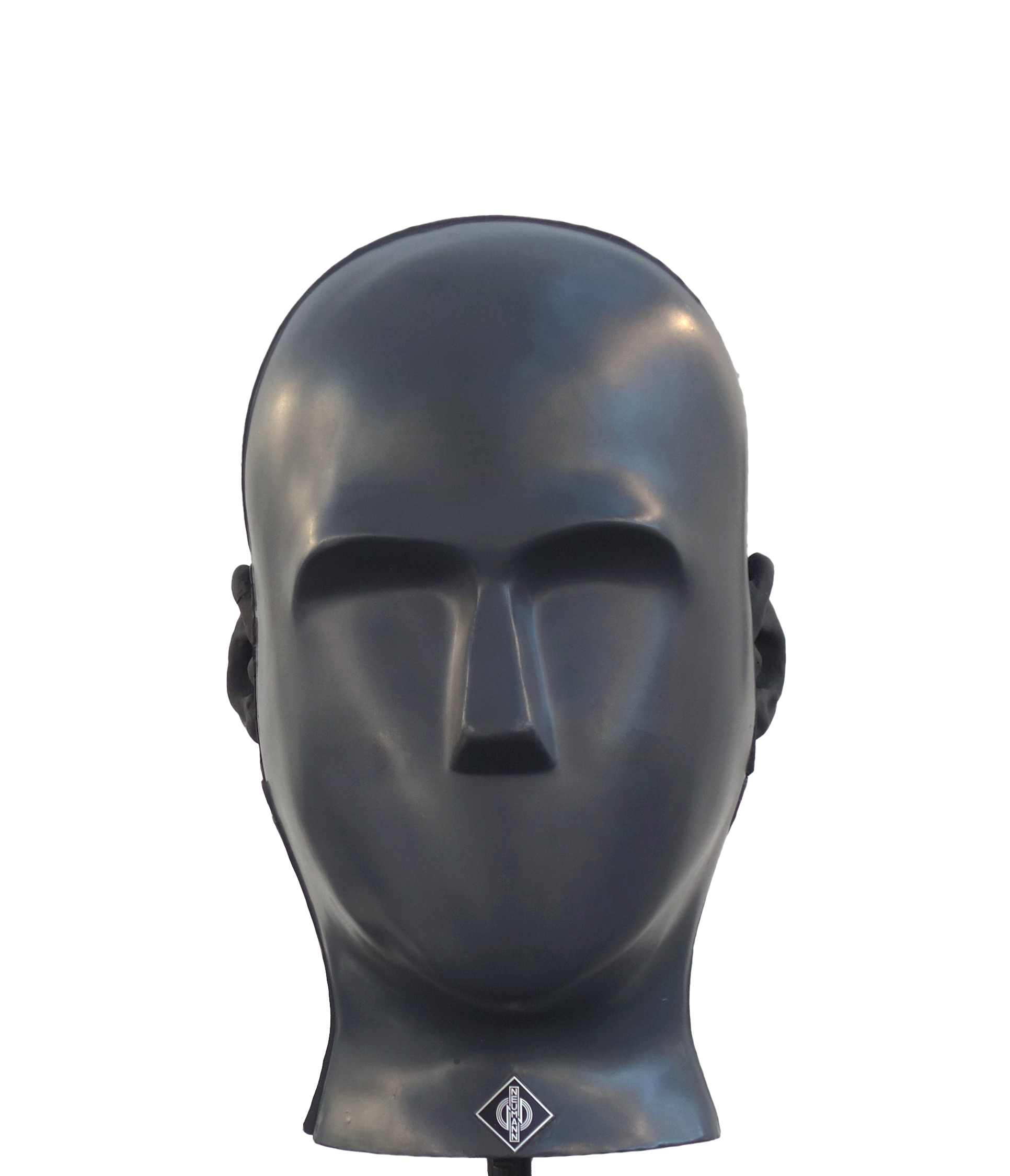}\\[-1mm]
        {\small RA/RV}
    \end{minipage}\hfill
    \begin{minipage}[t]{0.245\linewidth}
        \centering
        \includegraphics[width=\linewidth]{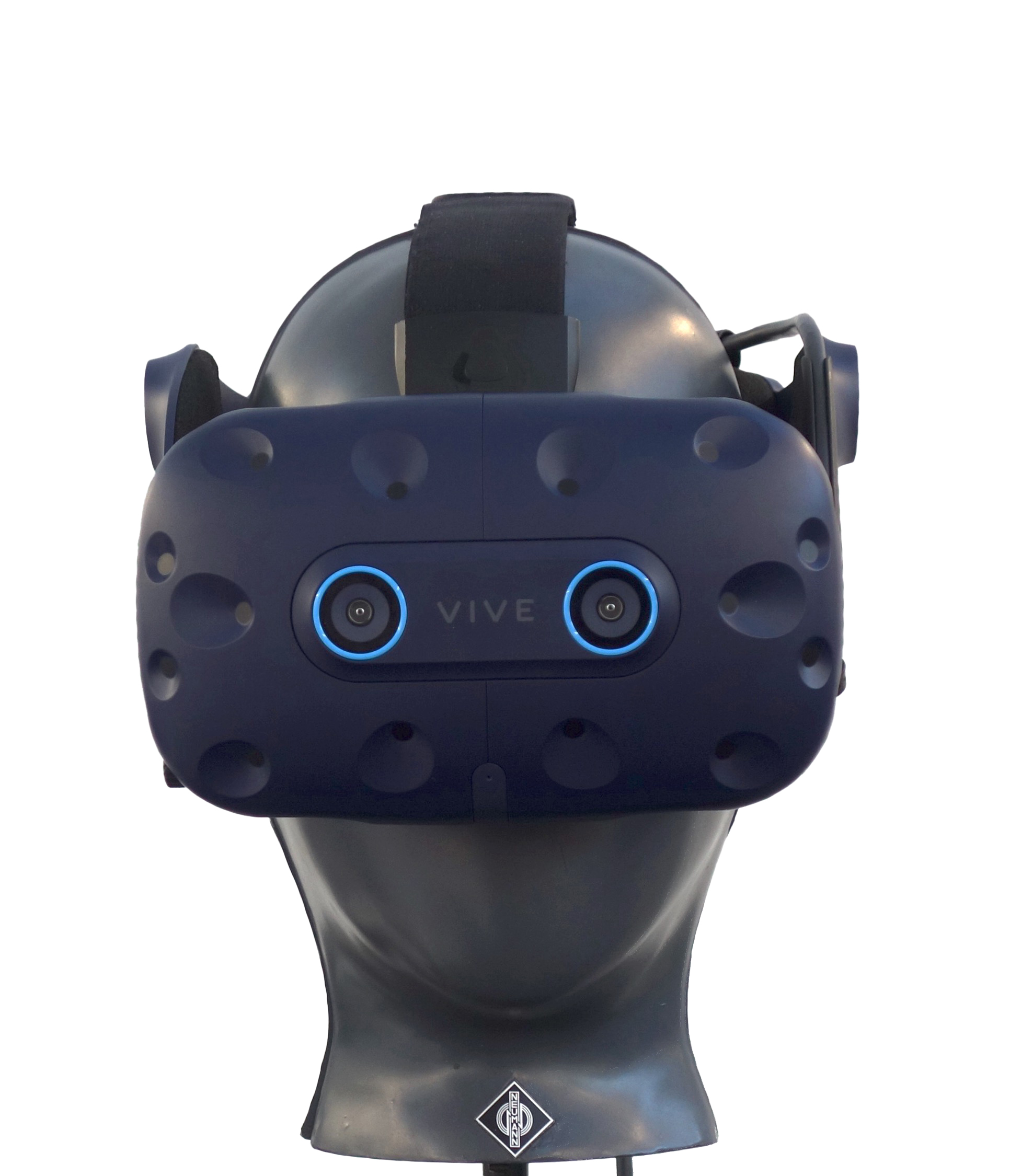}\\[-1mm]
        {\small RA/VV}
    \end{minipage}\hfill
    \begin{minipage}[t]{0.245\linewidth}
        \centering
        \includegraphics[width=\linewidth]{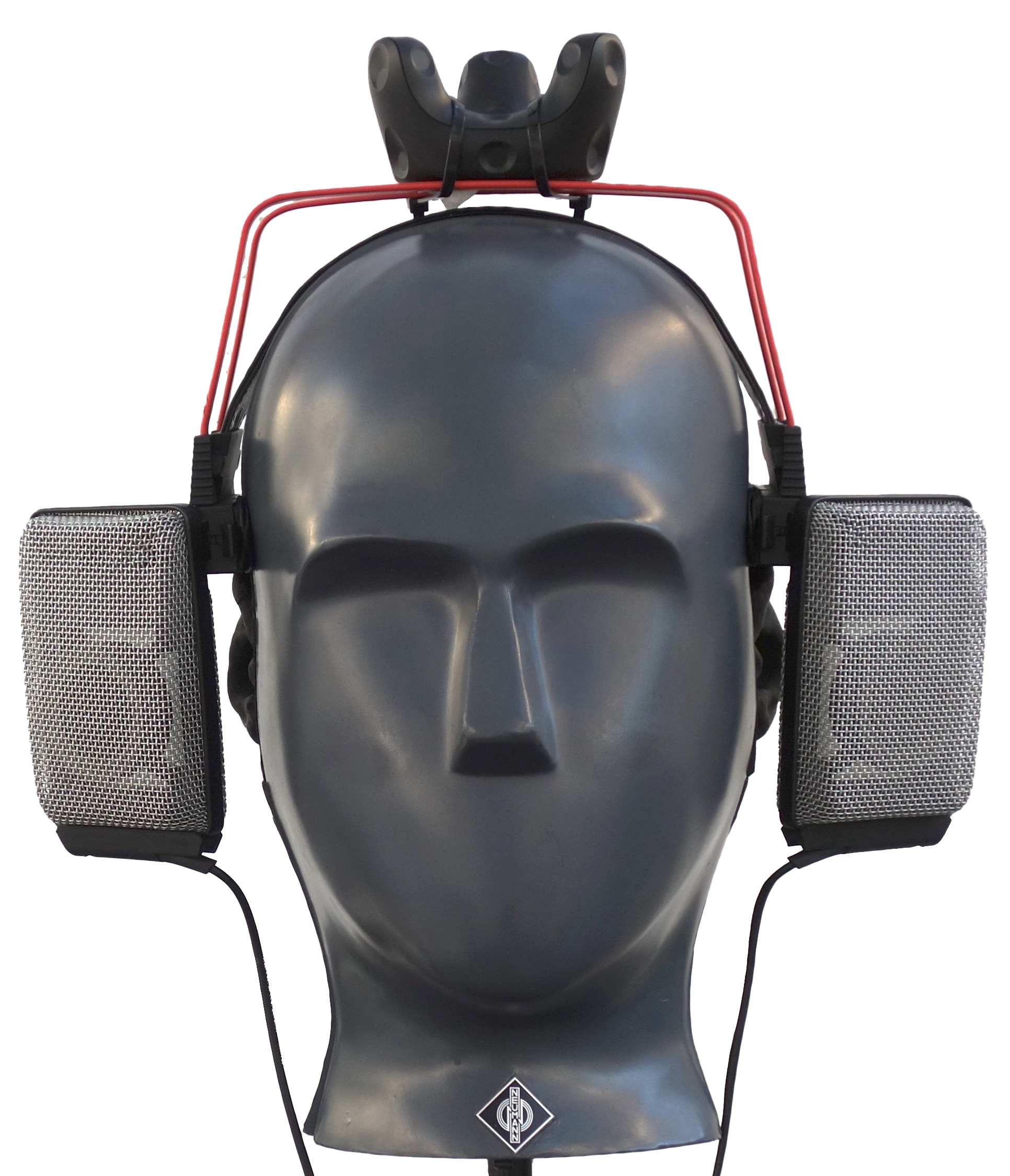}\\[-1mm]
        {\small AAR, VA/RV}
    \end{minipage}\hfill
    \begin{minipage}[t]{0.245\linewidth}
        \centering
        \includegraphics[width=\linewidth]{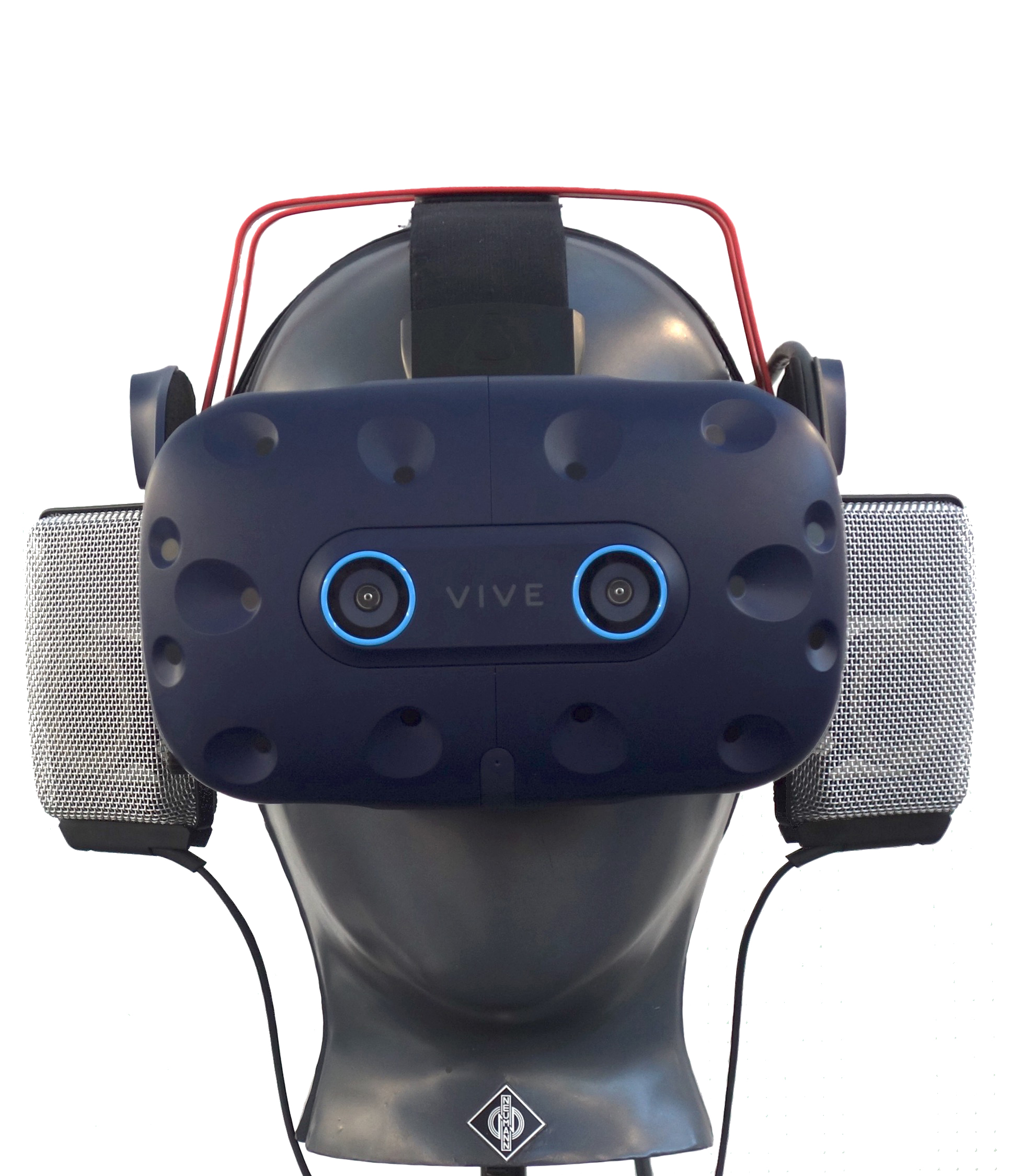}\\[-1mm]
        {\small VR, VA/VV}
    \end{minipage}

    \caption{Experimental conditions, depicted by the dummy head wearing the hardware used for auditory and visual virtualization. From left to right: real audio / real vision (RA/RV), real audio / virtual vision (RA/VV), audio-augmented reality (AAR) or virtual audio / real vision (VA/RV), and virtual reality (VR) or virtual audio / virtual vision (VA/VV).}
    \label{fig:virtualizationConditions}
\end{figure*}

The plausibility experiment had the two conditions AAR and VR, meaning vision was either real or virtualized. The ADP experiment had four conditions: real audio / virtual vision (RA/VV), virtual audio / virtual vision (VA/VV), real audio / real vision (RA/RV), and virtual audio / real vision (VA/RV). For conditions involving virtualized vision, participants wore the HMD that provided the real-time visual rendering of the room. For conditions involving virtual audio, participants wore the headphones providing the binaural rendering of the real-time room acoustic simulation. To provide a better overview and make the described conditions easier to remember, they are pictured in Fig.~\ref{fig:virtualizationConditions}. Please note that, for consistency, we used the notation VA/RV and VA/VV in the distance perception experiment; in terms of employed hardware, these correspond to the AAR and VR conditions, respectively, as denoted in the plausibility experiment.

\section{Plausibility Experiment}\label{sec:Exp1}
This experiment assessed the plausibility of the 6DoF XR audio system in AAR and VR contexts, following the plausibility paradigm proposed by Lindau and Weinzierl~\cite{Lindau2012}. The experiment evaluated the system's overall quality and provided new insights into how visual virtualization affects plausibility, thereby assessing whether spatial audio rendering has different requirements in AAR versus VR.

\subsection{Method}
\subsubsection{Stimuli}
In line with the concept of a plausibility test to evaluate the system's overall quality, we used a variety of audio content. More specifically, we used 13 anechoic mono audio files between 8 and \SI{12}{\second} long, including recordings of male and female speakers and singers, musical instruments such as a guitar, cello, and clarinet, as well as electronic music samples such as a drum loop. The audio files were loudness-normalized according to ITU-R BS.1770-4 to match the loudness of the pink noise test signal used to adjust the system's playback level. 

\subsubsection{Procedure}
In the experiment, participants were presented with a real or virtual sound source and had to answer the question, ``Was this a virtual loudspeaker?'' with ``yes'' or ``no''. Each participant completed 104 trials in total, consisting of all 13 test signals presented once through each of the eight loudspeakers. Half of the trials were presented via real loudspeakers and half were virtualized. All participants were presented with the same full set of signal-loudspeaker combinations, but the assignment of real versus virtual presentation and the order of trials were randomized individually. Additionally, to render small remaining level cues ineffective, the playback level of each trial was randomly roved within a range of $\pm$\SI{3}{\decibel} in \SI{1}{\decibel} increments.

To let participants experience the 6DoF audio and visual rendering, they were encouraged to move freely within the listening area while the stimuli were presented. This often led participants to kneel down toward LS8, get really close to LS1, turn around quickly, or cross the entire listening area, all to maximize cues for deciding whether the sound source was real or virtual. In the AAR condition, participants submitted their answers using the tablet they held, which displayed a simple TouchOSC GUI with the question and the two buttons ``yes'' or ``no''. In the VR condition, participants provided their answer by using the interactive menu displayed in the VR scene with the Vive controller. Submitting a response stopped playback of the current trial, if it was still playing, and started playback of the next trial after one second of silence. Participants could neither repeat a trial nor proceed without answering, and no feedback was provided. The video in the supplementary material~\cite{supp_material} illustrates the described test procedure and provides an impression of the experiment.

Before the experiment, the participants received instructions on the procedure and were equipped with the XR hardware specific to the experimental condition. Then, they completed a brief training session consisting of eight trials to become familiar with the setup and test procedure. As in the actual experiment, they did not receive feedback on their answers in the training session. A complete experimental session lasted about \SI{25}{$\min$}, including the verbal instructions and training.

\subsubsection{Data Analysis}
Following the standard analysis of SDT, the sensitivity $d'$, the corresponding 2AFC detection rate $p_{\mathrm{2AFC}}$, and the bias $\beta$, were calculated for each participant based on their yes/no responses and their respective hit and false alarm rates~\cite{Lindau2012,hautus_detection_2022}. Sensitivity $d'$ quantifies the perceived sensory difference between real and virtual audio reproduction, whereas the bias $\beta$ reflects the participant’s response tendency, independent of sensitivity, toward assuming that the sound source was real or virtual. The SDT parameters were estimated using log-linear transformed data~\cite{hautus_detection_2022}, preventing the rates from reaching 0 or 1, which would yield an infinite $d'$. In the present data, a hit rate of 0 appeared for one participant in the VR condition.

For statistical analysis, we first performed a Shapiro-Wilk test for normality, which indicated no violations of the normality assumption for the $d'$ values, but did indicate violations for the $\beta$ values. However, as parametric tests are generally robust to slight violations of the normality assumption~\cite{Glass1972}, we analyzed all the data using $t$-tests. In particular, we applied paired-samples $t$-tests to compare the results across conditions, as well as one-sample one-tailed $t$-tests to compare the data against different thresholds, all at a 0.05 significance level. Additionally, we performed Bayesian one-sample $t$-tests (JZS scaling factor $r$ = .707) for specific conditions, providing Bayes Factors (BF)---which, in contrast to common null-hypothesis significance testing, allow stating evidence in favor of the null hypothesis~\cite{Rouder2009}. All statistical analyses were conducted in R using \textit{jamovi} (Version 2.7.12) and \textit{RStudio} (Version 2025.09.2+418). Anonymized results and analysis files are provided in the supplementary material~\cite{supp_material}.

\subsection{Results}

\begin{figure*}[htb]
    \centering
    \begin{minipage}[h]{0.495\textwidth}
        \centering
        \includegraphics[width=\textwidth]{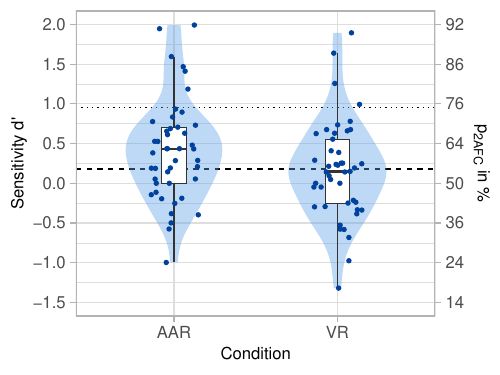}
    \end{minipage}
    \hfill
    \begin{minipage}[h]{0.495\textwidth}
        \centering
        \includegraphics[width=\textwidth]{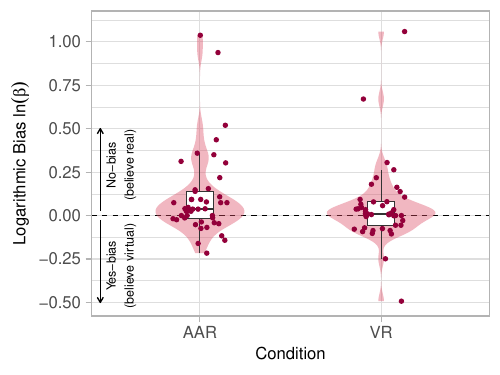}
    \end{minipage}
    \caption{Results of the plausibility experiment. The left plot shows sensitivity $d'$ and the right plot shows logarithmic bias $ln(\beta)$, each displayed for the two virtualization conditions, AAR and VR. In the sensitivity plot, the dashed horizontal line indicates the threshold of $d' = 0.1777$, and the dotted horizontal line indicates the threshold of $d' = 0.9539$. In both plots, dots represent individual values, box plots indicate the median and interquartile range (IQR), with whiskers displaying 1.5 $\times$ IQR below the 25th percentile or above the 75th percentile, and violins depict the full distribution shape and density of the data as a kernel density estimate, where wider sections indicate more common values.}
    \label{fig:PlausibilityResults}
\end{figure*}

Figure~\ref{fig:PlausibilityResults} shows the results of the experiment, with the left plot displaying the sensitivity $d'$ and the right plot displaying the logarithmic bias $ln(\beta)$, both with respect to the two virtualization conditions, AAR and VR. The sensitivity plot on the left reveals that a few participants obtained comparably high $d'$ values, indicating they could clearly differentiate between real and virtual sound sources. However, most participants obtained comparably low $d'$ values, suggesting differentiation was not easy, with an interquartile range (IQR) of $0.00$ to $0.71$ for the AAR condition and $-0.25$ to $0.56$ for the VR condition. The plot also suggests a higher degree of plausibility for the VR condition, with a mean value of $0.16$, compared to the AAR condition, which yielded a mean $d'$ value of $0.41$. A paired-samples $t$-test comparing both conditions confirmed that the VR condition yielded significantly lower sensitivity than the AAR condition [$t(44) = 2.27, p = .028, d = 0.34$], suggesting a higher degree of plausibility of the 6DoF spatial audio rendering in VR---that is, with visual virtualization. 

To determine if the rendering can be considered plausible, the sensitivity values can be compared to different thresholds. A one-sided $t$-test comparing whether the $d'$ values are significantly smaller than the more lenient threshold of $d' = 0.9539$ ($p_\text{2AFC}=0.75$)~\cite{Brinkmann2019a} indicates that both the AAR and VR systems are plausible [$t(44) = 5.69, p < .001, d = 0.85$ for AAR; $t(44) = 8.56, p < .001, d = 1.28$ for VR]. On the other hand, performing a one-sided $t$-test to determine if the d' values are significantly smaller than the more conservative threshold of $d' = 0.1777$ ($p_\text{2AFC}=0.55$)~\cite{Lindau2012} suggests that neither system can be considered plausible [$t(44) = 2.47, p = .991, d = 0.37$ for AAR; $t(44) = 0.17, p = .434, d = 0.03$ for VR]. However, the mean $d'$ of the VR condition is slightly below this threshold. A Bayesian one-sample $t$-test yielded $BF_{01} = 6.11$, providing moderate evidence in favor of the null hypothesis that the mean $d'$ of the VR condition does not differ from the threshold of $d' = 0.1777$, indicating that it does not exceed this threshold. This can be interpreted as evidence for the VR condition being plausible, even when the strictest threshold is applied.

The bias plot in Fig.~\ref{fig:PlausibilityResults} (right) shows participants' response tendencies, independent of actual sensory differences. Values of $ln(\beta) > 0$ indicate a tendency to answer ``No,'' implying they tended to believe the stimuli were real, while values of $ln(\beta) < 0$ indicate a tendency to answer ``Yes,'' implying they tended to assume the stimuli were virtual. In both conditions, few participants obtained high values of $ln(\beta)\approx1$, suggesting they strongly believed in the realness of the stimuli. Notably, the participants did not exhibit equally pronounced ``Yes'' biases. However, most participants obtained $ln(\beta)$ values close to $0$, with a mean $ln(\beta)$ value of $0.11$ and an IQR ranging from $-0.02$ to $0.14$ for the AAR condition, as well as $0.01$ and an IQR ranging from $-0.06$ to $0.08$ for the VR condition. These results indicate either no clear response tendency or a slight tendency to answer ``No'' in case of the AAR condition. A one-sided $t$-test confirmed that for the AAR condition, the $ln(\beta)$ values are significantly higher than $0$ [$t(44) = 2.95, p = .003, d = 0.44$ for AAR; $t(44) = 0.14, p = .444, d = 0.02$ for VR], revealing that, on average, participants tended to believe in the realness of the stimuli slightly more when assessing the 6DoF spatial audio rendering in AAR---that is, with real vision.

\subsection{Discussion}\label{sec:Exp1Discussion}
The experiment demonstrates a high degree of plausibility of the 6DoF XR audio system. Most participants exhibited low sensitivity values, which were clearly below the lenient threshold and, for the VR condition, even within the range of the more conservative threshold. This aligns with the participants' subjective reports, as the majority stated that they were mostly guessing. 

Notably, three participants showed negative sensitivities of $d'\lesssim -1$ ($p_\text{2AFC}\lesssim 0.25$), indicating that they systematically confused simulated and real stimuli. In a different setting, these participants might have been able to reliably detect the simulation; for example, if they had received feedback during training, or if the test had used more familiar audio content.

The bias results provide additional insight into listeners’ decision-making strategies. Although the $ln(\beta)$ values were generally close to $0$ in both conditions, indicating no strong systematic response tendency, the AAR condition showed a small but significant shift toward answering ``No,'' meaning participants leaned more toward believing the stimuli were real. This bias was likely expectation-driven: seeing the real loudspeakers may have predisposed listeners to assume real audio reproduction, even though the sensitivity data demonstrate that they were, in fact, better able to detect differences in AAR than in VR.

Comparing AAR and VR reveals that visual virtualization affects the plausibility of XR audio reproduction. Sensitivity was significantly lower in VR, suggesting that introducing a virtual visual environment reduced participants’ ability to distinguish between real and virtual sound sources. In AAR, by contrast, access to the real visual scene appears to have supported slightly more accurate discrimination. These results demonstrate that while both AAR and VR can create plausible auditory experiences, the visual context affects how well participants can detect differences between real and virtual audio presentations.

\section{Dynamic Auditory Distance Perception Experiment}\label{sec:Exp2}
This experiment assessed dynamic ADP under four different virtualization conditions (RA/RV, RA/VV, VA/RV, and VA/VV; see Figs.~\ref{fig:procedure} and~\ref{fig:virtualizationConditions}) to systematically investigate the effects of modality-specific and combined virtualization on dynamic ADP in a locomotion-based distance estimation task, offering new insights into how cross-modal effects impact fundamental perceptual processes, such as ADP.

\subsection{Method}
\subsubsection{Stimuli}
We used the first four phonetically balanced sentences from the first list of Harvard sentences, spoken by a native female British English speaker~\cite{demonte_2019}, as anechoic test signals. The four sentences were arranged in sequence with \SI{0.25}{\second} pauses between each one, for a total signal length of \SI{10}{\second}. The final sequence was loudness-normalized according to ITU-R BS.1770-4 to match the loudness of the pink noise signal used to adjust the overall playback level. 

\subsubsection{Procedure}
In the experiment, the speech signal was randomly played from one of the five loudspeakers (LS1 to LS5) arranged along the line. Depending on the virtualization condition, the sound and visual representations were either real or virtual. In each trial, participants had to walk up and down the walking line during presentation of the speech signal, whereby they had to begin at the end of the listening area, walk toward the loudspeakers as soon as the audio started, turn around in front of LS1, and then return to the starting position. This took approximately \SI{10}{\second}, allowing participants to hear the speech signal throughout their walk, with the stimulus completing when they returned to the starting position. Once back at the starting position, participants were asked to turn to face the loudspeakers and report which loudspeaker had played the sound by selecting a number between one and five. An experimental session consisted of 12 blocks, each containing one presentation of all five loudspeakers in randomized order, resulting in 12 trials per loudspeaker and a total of 60 trials per condition. Participants could not repeat a trial or proceed without answering, and no feedback was provided. Submitting a response started playback of the next trial after a one-second silence period.

In real-vision sessions, i.e., when participants did not wear the HMD, they submitted their answers using the tablet they held. The tablet displayed a simple TouchOSC GUI with the question, ``Which loudspeaker played the sound?'' and five buttons numbered one to five, arranged vertically to match the order of the loudspeakers. In virtual-vision sessions, i.e., when participants wore the HMD, they provided their answers using the interactive menu displayed in the VR scene with the Vive controller. Similar to the tablet's GUI, the VR menu displayed the question and five vertically arranged buttons with the numbers one to five. The same numbers were provided on cardboard panels in the real and virtual visual scene (cf. Fig.~\ref{fig:MRL_Real_Unreal}). The video in the supplementary material~\cite{supp_material} illustrates the test procedure described here and provides an impression of the experiment.

Since participants were only allowed to walk along the walking line, their movement was restricted to the loudspeaker axis, yielding dynamic ADP cues such as changes in sound level and DRR (and, to some extent, dynamic binaural cues when turning), while largely eliminating motion parallax cues, i.e., changes in source direction that arise when moving off-axis, which could aid source identification~\cite{Zahorik2005,Kolarik2016}. 
This restriction was a deliberate methodological choice to focus the locomotion-based experiment on the two dynamic ADP cues of interest---level and DRR---while minimizing additional directional cues that support source identification rather than source distance estimation.

Before each experimental session, participants were instructed on the procedure and, if applicable, equipped with the XR hardware specific to the respective virtualization condition. Prior to each of the four sessions, they completed a short training phase consisting of ten trials with a fixed sequence: the first five trials presented LS1 to LS5 in ascending order, and the subsequent five trials presented LS5 to LS1 in descending order. Participants were informed about this fixed sequence, and the training included the full walking and response procedure, providing both procedural familiarization and limited perceptual practice with the dynamic ADP cues available in that condition. 

After each session, participants were asked to rate, on a scale from one to thirteen, how much effort was required to estimate the auditory distance of the loudspeakers. We used an established listening effort measurement scale~\cite{luts_multicenter_2010}, with seven of the in total thirteen categories labeled with effort scaling categorical units, namely, no effort (1), very little effort (3), little effort (5), moderate effort (7), considerable effort (9), much effort (11), and extreme effort (13). This measure was included in the experiment to assess whether virtualization affects distance estimation effort. A complete experimental session lasted about \SI{15}{$\min$}, including verbal instructions and training trials.

\subsubsection{Data Analysis}
The data from the loudspeaker identification task were analyzed using complementary approaches. For descriptive analysis, the raw responses (i.e., the reported loudspeaker number per trial) were summarized as response proportions for each combination of target and response loudspeaker, separately for each virtualization condition. For inferential statistical analysis, two dependent variables were derived from the trial-level responses: \textit{Accuracy}, coded as a binary variable (1 for a correct response, 0 otherwise), and signed \textit{Error}, computed as the difference between the response and the target loudspeaker numbers (negative values indicating a bias toward nearer loudspeakers, positive values toward farther loudspeakers). Accuracy was analyzed using a generalized linear mixed-effects model (GLMM), and error was analyzed using a linear mixed-effects model (LMM). For consistency in statistical methodology across outcome measures, the distance estimation effort ratings, hereafter \textit{Effort}, were also analyzed using an LMM. All statistical analyses were conducted in R using jamovi (Version 2.7.12) and RStudio (Version 2025.09.2+418) in combination with the GAMLj3 suite (Version 3.6.5,~\cite{githubGamlj}) for the linear models. Anonymized response data and analysis files are provided in the supplementary material~\cite{supp_material}.

The GLMM for accuracy used a binomial distribution with a logit link function (logistic model) to account for the binary dependent variable. The model was estimated using maximum likelihood. The fixed-effects structure included \textit{Condition}, \textit{Loudspeaker}, and their interaction, while the random-effects structure comprised by-participant random intercepts and random slopes for \textit{Condition} to account for individual differences in baseline performance and condition-specific effects:
\begin{equation}
\begin{aligned}
\mathrm{Accuracy} \sim {} & 
\mathrm{Condition} + \mathrm{Loudspeaker} \\
& + \mathrm{Condition}:\mathrm{Loudspeaker} \\
& + (1 + \mathrm{Condition} \mid \mathrm{Participant}).
\end{aligned}
\label{eq:glm_accuracy}
\end{equation}

The LMM for signed error used restricted maximum likelihood (REML) estimation and Satterthwaite's method to obtain the degrees of freedom for inferring the fixed effects. The model employed the same fixed- and random-effects structure as the GLMM; that is, the model formula was identical to Eq.~(\ref{eq:glm_accuracy}), except that \textit{Error} replaced \textit{Accuracy} as the dependent variable. Visual inspection of residual diagnostics (residual histogram and Q-Q plot) revealed acceptable normality and homoscedasticity. For both the accuracy GLMM and the signed-error LMM, categorical predictors were coded using treatment (dummy) coding for \textit{Condition} (RA/RV as reference) and orthogonal polynomial contrasts for \textit{Loudspeaker} to capture linear and higher-order trends across the ordered distance positions.

The LMM for the distance estimation effort ratings also employed REML estimation and Satterthwaite's method. The model had a much simpler structure, with the fixed effect \textit{Condition} and by-participant random intercepts, yielding
\begin{equation}
\begin{aligned}
\mathrm{Effort} \sim {} & 
\mathrm{Condition} + (1 \mid \mathrm{Participant}).
\end{aligned}
\end{equation}
Residual diagnostics indicated normality and homoscedasticity.

Post-hoc comparisons for significant fixed effects were conducted on the model-based estimated marginal means, and adjusted using Holm’s method. Holm-corrected p-values are reported throughout.

\subsection{Results}

\subsubsection{Confusion Matrices}

\begin{figure*}[h]
    \centering
    \includegraphics[width=1\textwidth]{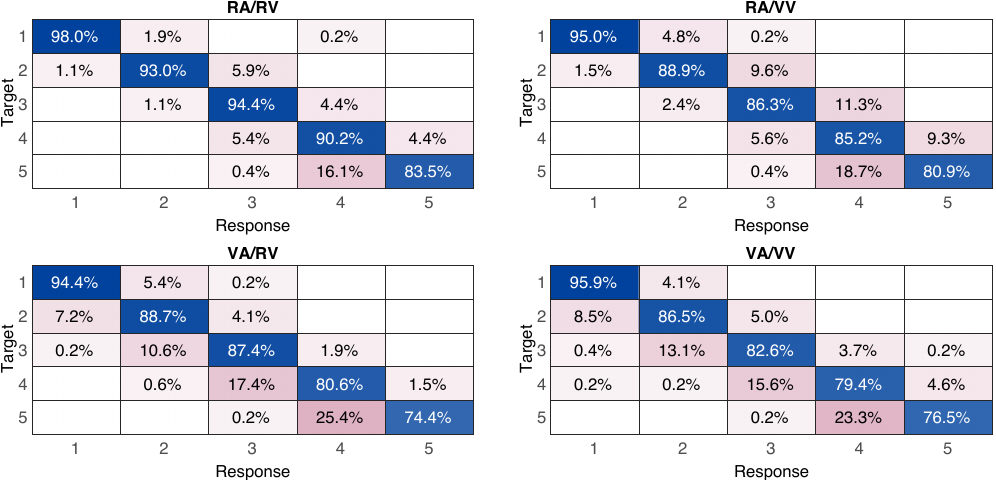}
    \caption{Confusion matrices showing the response proportions across all participants as a function of the response loudspeaker and target loudspeaker (x- and y-axes, respectively) for the four virtualization conditions.}
    \label{fig:confusion_matrix}
\end{figure*}

The confusion matrices in Fig.~\ref{fig:confusion_matrix} show the response proportions across all participants as a function of the response loudspeaker and target loudspeaker for each of the four virtualization conditions. Diagonal entries represent correct loudspeaker identifications and are analyzed statistically using the GLMM for accuracy. Off-diagonal entries indicate misclassifications and reveal systematic response trends, such as biases toward nearer or farther loudspeakers, which are examined statistically using the LMM for the signed error. The plots reveal systematic response patterns across virtualization conditions. In general, the proportion correct decreases with increasing loudspeaker distance, yielding the highest accuracy for LS1 and the lowest for LS5; the only minor deviation from this trend occurs in RA/RV, where accuracy for LS3 is slightly higher than for LS2. Participants' performance was best in the fully real condition (RA/RV) and decreased with any form of virtualization. Nevertheless, performance remained comparatively high across all virtualized conditions, with proportions correct ranging from about 74\% (LS5, VA/RV) up to about 96\% (LS1, VA/VV). The off-diagonal entries reveal that misclassifications were largely confined to adjacent loudspeakers, indicating that ordinal distance information was generally preserved, even when absolute identification accuracy declined. For LS1 and LS5, misclassifications consistently occurred toward the farther (LS2) or nearer (LS4) loudspeaker, respectively, reflecting boundary effects due to the absence of a closer or more distant response option. For LS2 to LS4, local errors differ in direction across conditions: responses tend to be biased toward farther loudspeakers in real audio conditions (RA/RV, RA/VV) and toward nearer loudspeakers in the virtual audio conditions (VA/RV, VA/VV). 

\subsubsection{Accuracy}

\begin{figure*}[h]
    \centering
    \includegraphics[width=1\textwidth]{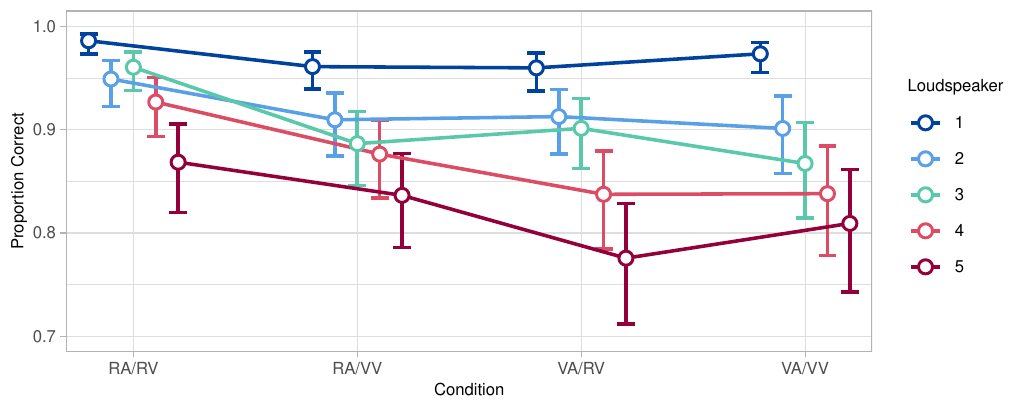}
    \caption{Estimated marginal means of trial-level accuracy obtained from the GLMM as a function of virtualization condition (x-axis) and loudspeaker (color-coded). Error bars indicate 95\% confidence intervals.}
    \label{fig:emm_pCorr_condition_loudspeaker}
\end{figure*}

In line with these observations, the GLMM revealed strong main effects of both condition [$\chi^2(3) = 27.29, p < .001$] and loudspeaker [$\chi^2(4) = 78.02, p < .001$] on trial-level accuracy, as well as a modest but significant interaction [$\chi^2(12) = 22.28, p = .035$], which indicates that the effect of loudspeaker distance differed slightly across virtualization conditions. The estimated marginal means in Fig.~\ref{fig:emm_pCorr_condition_loudspeaker} clearly reflect these effects. The fully real condition (RA/RV) yielded the highest proportion correct, whereas all virtualized conditions showed reduced performance. Furthermore, accuracy decreased with increasing loudspeaker distance, with a single exception (LS3 in RA/RV). The rate of this decrease varied slightly between conditions, yielding the significant interaction effect.
 
The estimated model parameters provide further detail on these patterns. Relative to the fully real condition RA/RV, all virtualized conditions showed significantly lower trial-level accuracy, reflected in negative coefficient estimates with odds ratios (ORs) below 1 [RA/VV: estimate $= -0.73$, SE $= 0.18$, OR $= 0.48$, $p < .001$; VA/RV: estimate $= -0.84$, SE $= 0.18$, OR $= 0.43$, $p < .001$; VA/VV: estimate $= -0.80$, SE $= 0.24$, OR $= 0.45$, $p < .001$]. These values indicate that the odds of a correct response were reduced by about half in all virtualized conditions compared to the fully real baseline. The effect of loudspeaker showed a strong linear trend (estimate $= -1.62$, SE $= 0.22$, OR $= 0.20$, $p < .001$), confirming that accuracy declined markedly with increasing loudspeaker distance. Significant cubic ($p = .004$) and quartic ($p = .020$) polynomial components further indicate that this decrease was not purely linear. 

\begin{figure}[h]
    \centering
    \includegraphics[width=.5\textwidth]{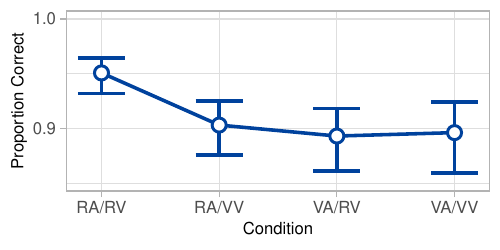}
    \caption{Estimated marginal means of trial-level accuracy obtained from the GLMM for the four virtualization conditions, pooled across loudspeakers. Error bars indicate 95\% confidence intervals.}
    \label{fig:emm_pCorr_condition}
\end{figure}

To further focus on the effect of virtualization, Fig.~\ref{fig:emm_pCorr_condition} shows the estimated marginal means of trial-level accuracy for each virtualization condition, pooled across loudspeakers. Consistent with the previous analysis, accuracy was highest for the RA/RV condition (M $= 0.95$) and decreased for all conditions involving virtualization (RA/VV: M $= 0.90$, VA/RV: M $= 0.89$, VA/VV: M $= 0.90$). Post-hoc comparisons confirmed that RA/RV differed significantly from each of the three virtualized conditions (all $p < .003$), whereas no significant differences occurred among the conditions with virtualization themselves (all $p = 1$). 

\subsubsection{Error}

\begin{figure*}[h]
    \centering
    \includegraphics[width=1\textwidth]{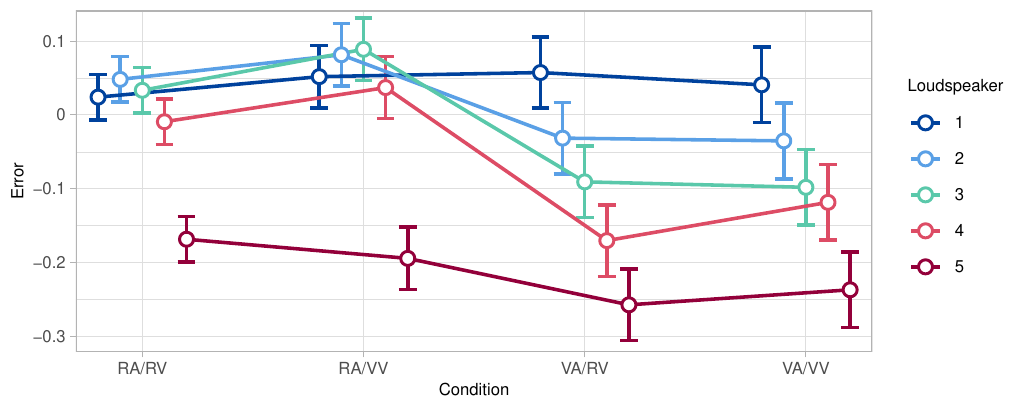}
    \caption{Estimated marginal means of trial-level signed error obtained from the LMM as a function of virtualization condition (x-axis) and loudspeaker (color-coded). Error bars indicate 95\% confidence intervals.}
    \label{fig:emm_error_condition_loudspeaker}
\end{figure*}

The signed-error analysis provided complementary insights into the nature of distance judgment errors beyond overall accuracy. The LMM revealed the same pattern of significant effects for signed error as the GLMM did for accuracy: main effects of condition [$F(3,44) = 11.60$, $p < .001$] and loudspeaker [$F(4,10604) = 205.79$, $p < .001$], as well as a significant interaction [$F(12,10604) = 9.67$, $p < .001$], indicating that the magnitude and direction of the signed error, and hence also the response bias toward nearer or farther loudspeakers, varied across loudspeakers and virtualizations. 

The estimated marginal means in Fig.~\ref{fig:emm_error_condition_loudspeaker} illustrate these effects. In the fully real condition (RA/RV), signed errors are close to $0$ for LS1 to LS4, indicating mostly unbiased distance judgments, whereas LS5 shows a pronounced negative error. This pattern reflects a boundary effect evident across all virtualization conditions, as responses for LS5 could only be biased toward the nearer distance due to the absence of a farther distance option. Similarly, LS1 exhibits slightly positive errors across all virtualization conditions, reflecting the inverse boundary effect. A similar pattern appears in the RA/VV condition, though with slightly higher absolute error values, indicating a trend of stronger distance overestimation for LS1 to LS4 and stronger underestimation for LS5. In contrast, both conditions involving virtual audio (VA/RV and VA/VV) exhibit systematic negative signed errors that become more pronounced with loudspeaker distance, indicating a bias toward nearer loudspeakers. This compression of perceived distance was most pronounced for the more distant loudspeakers.

The fixed-effect estimates confirm these patterns. Relative to RA/RV, the mean signed error was significantly more negative in VA/RV (estimate $= -0.084$, SE $= 0.02$, $p < .001$) and VA/VV (estimate $= -0.075$, SE $= 0.02$, $p = .002$), indicating systematic underestimation of source distance with auditory virtualization. The observed positive shift for RA/VV was not significant (estimate $= 0.027$, SE $= 0.02$, $p = .134$), though the effect became stronger when the boundary conditions LS1 and LS5 were excluded (estimate $= 0.045$, SE $= 0.02$, $p = .053$), essentially rendering this observation a trend rather than a clear effect. Across conditions, the signed error became increasingly negative as the loudspeaker distance increased, with a negative quadratic term indicating that this underestimation grew progressively steeper at larger distances (linear: estimate $= -0.140$, SE $= 0.01$, $p < .001$; quadratic: estimate $= -0.105$, SE $= 0.01$, $p < .001$). This effect was most pronounced in VA/RV and VA/VV.

\begin{figure}[h]
    \centering
    \includegraphics[width=.5\textwidth]{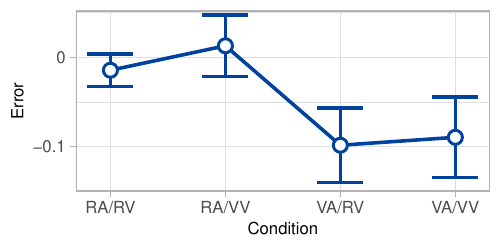}
    \caption{Estimated marginal means of trial-level signed error obtained from the LMM for the four virtualization conditions, pooled across loudspeakers. Error bars indicate 95\% confidence intervals.}
    \label{fig:emm_error_condition}
\end{figure}

Figure~\ref{fig:emm_error_condition} further highlights the effect of virtualization by showing the estimated marginal means of the trial-level signed error for each virtualization condition, pooled across loudspeakers. Post-hoc comparisons confirmed that VA/RV (M $= -0.099$) and VA/VV (M $= -0.090$) resulted in significantly larger negative biases than RA/RV (M $= -0.014$), all with $p \leq .007$. Meanwhile, RA/VV exhibited a slight positive bias (M $= 0.013$), though it did not differ significantly from the fully real condition ($p = .268$). No significant difference was found between VA/RV and VA/VV ($p = .765$).

\subsubsection{Distance Estimation Effort}

\begin{figure}[h]
    \centering
    \includegraphics[width=.5\textwidth]{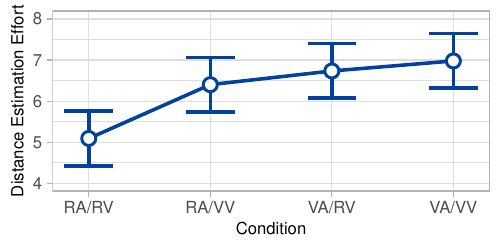}
    \caption{Estimated marginal means of distance estimation effort obtained from the LMM for the four virtualization conditions. Error bars indicate 95\% confidence intervals. Effort was rated on a 13-point categorical scale from 1 (“no effort”) to 13 (“extreme effort”).}
    \label{fig:emm_effort_condition}
\end{figure}

The LMM used to analyze the distance estimation effort ratings revealed a significant main effect of condition [$F(3,132) = 11.95$, $p < .001$], indicating that virtualization systematically increased the required effort. The estimated marginal means, shown in Fig.~\ref{fig:emm_effort_condition}, demonstrate that effort was lowest in the fully real condition RA/RV (M $= 5.09$), corresponding to an average rating close to “little effort”, whereas virtualized conditions required higher effort: RA/VV (M $= 6.40$), VA/RV (M $= 6.73$), and VA/VV (M $= 6.98$), all of which are close to the “moderate effort” category. Post-hoc comparisons confirmed that RA/RV differed significantly from each of the three virtualized conditions (all $p \leq .001$), whereas no significant differences were found among RA/VV, VA/RV, and VA/VV themselves (all $p \geq .286$). The supplementary material~\cite[Fig.\,S17]{supp_material} additionally provides violin plots of the effort ratings.

\subsection{Discussion}\label{sec:Exp2Discussion}

The dynamic ADP experiment revealed that virtualization systematically impaired distance estimation performance relative to the fully real baseline. Although performance remained relatively high, the three virtualized conditions (RA/VV, VA/RV, and VA/VV) yielded lower overall accuracy than the RA/RV condition, with a similar decrease across conditions. Correct responses decreased with increasing loudspeaker distance across all conditions, and most incorrect responses were confined to adjacent loudspeakers, indicating that ordinal distance perception was largely preserved. These results suggest a modality-independent cost of transitioning from a fully real setting to any virtualized condition in the examined dynamic ADP task.

The signed-error analysis revealed that auditory and visual virtualization affect distance judgments differently. When audio was virtualized (VA/RV, VA/VV), responses showed systematic, distance-dependent biases toward nearer loudspeakers, reflecting a compression of perceived distance that increased with source distance. In contrast, visual-only virtualization (RA/VV) did not produce a comparable directional bias; signed errors were small on average, with only a weak, non-significant trend toward a slight overestimation of distance. Thus, auditory virtualization primarily introduced a systematic bias toward nearer loudspeakers in perceived distance, whereas visual-only virtualization mainly reduced overall accuracy without introducing a significant bias toward nearer or farther loudspeakers.

The distance estimation effort ratings complemented these patterns. All virtualized conditions required significantly more effort than RA/RV, with mean ratings shifting from ``little effort'' to ``moderate effort''. In summary, these results demonstrate that, in the present 6DoF XR system, (i) any form of virtualization leads to a general decrease in performance and an increase in effort in dynamic ADP, and (ii) auditory virtualization leads to a systematic underestimation of source distance, while visual-only virtualization primarily increases uncertainty.

\section{General Discussion}

XR technology provides an opportunity to study spatial hearing as a multisensory perceptual process in controlled yet highly naturalistic conditions. However, its suitability for rigorous perceptual research depends on its perceptual fidelity and ability to elicit behavior comparable to that in the real world. In this study, we implemented a state-of-the-art 6DoF XR system that combines real-time room acoustic simulation and binaural rendering with a visually detailed virtual model of a large laboratory room. Using this system, we investigated plausibility and dynamic ADP in real and virtualized audio-visual conditions. Specifically, we addressed four research questions, examining whether the system provides plausible binaural rendering (RQ1), how visual virtualization affects plausibility (RQ2), how auditory and visual virtualization, separately or jointly, influence dynamic ADP as a fundamental perceptual process in a locomotion-based task (RQ3), and whether their effects combine additively in the fully virtual condition (RQ4).

\subsection{Plausibility and the Role of Visual Context}

Regarding RQ1, the plausibility experiment confirmed that the 6DoF XR audio system can provide plausible binaural rendering (see Sec.~\ref{sec:Exp1Discussion}). That most participants reported finding it difficult to distinguish between real and virtual sound sources is consistent with results from Himmelein and Pörschmann~\cite{Himmelein2026}, who showed that listeners can exhibit low confidence (type-II sensitivity) in their responses even when they are able to reliably detect the virtual sources.

These results extend previous work on the plausibility of binaural rendering to a real-time, simulation-based 6DoF XR system. Earlier plausibility studies primarily examined measurement-based AAR audio rendering with restricted listener movement~\cite{Lindau2012,Pike2014,Arend2021c,Neidhardt2021,lubeck_perceptual_2024,MeyerKahlen2024,Himmelein2025}. The present results demonstrate that real-time room acoustic simulation can achieve a high degree of plausibility despite relying on a shoebox approximation, non-individual HRTFs, and efficient simulation and binauralization rather than measured BRIRs. This is important because computationally efficient, adaptable real-time approaches are more practical for interactive XR applications than measurement-based solutions that usually require (sometimes rather dense) a priori measurements.

In response to RQ2, the plausibility experiment showed that the same XR audio rendering was perceived as more plausible in VR, when embedded in a virtual visual environment, than in AAR, when presented alongside the real visual scene (see Sec.~\ref{sec:Exp1Discussion}). This finding demonstrates that the visual context modulates the ability to detect subtle differences between real and virtual audio presentations. At the same time, participants in the AAR condition showed a slight yet significant tendency to judge the presented sound sources as real, likely because seeing the real loudspeakers predisposed them to assume real audio reproduction, even though they were better able to detect differences in the AAR condition than in the VR condition.

The difference between AAR and VR suggests that plausibility is not solely determined by audio rendering, but by the audiovisual reference frame in which the rendering is evaluated, clearly revealing cross-modal effects. In AAR, participants had access to real-world visual information about the room and loudspeaker locations, providing an accurate visual reference for determining whether the auditory event was consistent with the real environment. This likely enhanced their sensitivity to audio-visual mismatches and facilitated real-virtual audio discrimination. In VR, by contrast, virtualizing this visual reference likely weakened visual cues supporting accurate real-virtual audio discrimination. Thus, even though the virtual visual representation was detailed and visually coherent with the real room, it did not provide the same accurate reference; this likely reduced participants' sensitivity to audio-visual mismatches and increased the apparent plausibility of the binaural rendering.

Taken together, the results suggest that AAR and VR impose different perceptual requirements on XR audio rendering. In AAR, the real visual environment may demand more accurate audio rendering because mismatches can be detected relative to the real room, whereas VR appears to relax the perceptual constraints on audio rendering due to the virtual visual reference frame. These findings align with the results of an audiovisual matching study by Meyer-Kahlen et al.~\cite{meyer-kahlen_vision--audio_2025}, which also indicated different acoustic rendering requirements for VR and AAR due to different auditory reference frames, with VR generally having more relaxed requirements.

\subsection{Effects of Virtualization on Dynamic Auditory Distance Perception}

With respect to RQ3, the dynamic ADP experiment showed that any form of virtualization reduced accuracy and increased effort (see Sec.~\ref{sec:Exp2Discussion}). This suggests that, even when the virtual environment is well-matched to the real room, virtualization can increase uncertainty in a focused perceptual task. Nevertheless, performance remained high, with an average accuracy of about 90\% in the virtualized conditions compared to about 95\% in the fully real condition, and most errors confined to adjacent loudspeakers, suggesting that participants could still effectively utilize the available dynamic distance cues.

Across all conditions, accuracy decreased with increasing loudspeaker distance. This is consistent with the general properties of ADP, in which distance judgments become less precise for farther sound sources because distance-dependent changes in intensity and DRR become smaller, leading to a so-called acoustic horizon beyond which changes in distance are barely perceivable~\cite{Zahorik2005,Kolarik2016}. In our study, this effect was observed between loudspeakers, as well as across the walking line for one particular loudspeaker (see Fig.\,\ref{fig:BOL_DRR_RMS_RvsV}). 

In contrast to the similar decrease in accuracy across all virtualized conditions, the signed-error analysis revealed modality-specific effects. Auditory virtualization led to a systematic bias toward nearer loudspeakers that increased with distance, whereas visual-only virtualization did not cause a significant signed-error bias (see Sec.~\ref{sec:Exp2Discussion}). The bias in the virtual audio conditions aligns with the measured DRR differences, as virtual LS2 to LS5 generally led to slightly higher broadband DRR values than their real counterparts along the walking line (see Fig.\,\ref{fig:BOL_DRR_RMS_RvsV}), which corresponds to a closer perceived source and is therefore consistent with the observed distance underestimations. However, it is important to note that this occurred even though the simulated room was carefully matched to the real room (see Fig.\,\ref{fig:T20_DRR_LUFS}), with measured DRR differences not exceeding reported JNDs for the respective DRR ranges~\cite{Larsen2008}, and with only small level differences, mostly below \SI{1}{\decibel} (see Fig.\,\ref{fig:BOL_DRR_RMS_RvsV}). Thus, the observed bias is unlikely to reflect a gross mismatch in the simulated acoustics.

One possible explanation for the bias in the virtual audio conditions (VA/RV and VA/VV) is the combination of various deviations from real-world perception in the rendering. Such factors might be the use of non-individual HRTFs, which can increase localization blur, though their role in distance estimation is not entirely clear~\cite{Best2020,Arend2021b}; the shoebox approximation of the geometry, which can lead to slightly different early reflection patterns; or possible minor artifacts in the dynamic rendering in general. Together, these effects may lead to subtle deviations that affect the dynamic ADP task. Additionally, conventional static room acoustic metrics that are important for distance perception, such as DRR JNDs~\cite{Larsen2008}, may not fully predict performance in dynamic ADP. In the present locomotion-based task, listeners continuously integrated changes in level, DRR, binaural cues, and self-motion-related information. Small deviations that are near or below threshold in isolated, static comparisons may become relevant when they are present in a dynamic scene, or when they interact with aforementioned rendering approximations. Last, the broadband DRR comparison may not accurately reflect real-world perceptual DRR sensitivity. The supplementary frequency-dependent analysis showed that DRR differences were partly driven by octave bands from \SI{4}{\kHz} upwards. Because frequency-specific DRR JNDs are not known, we cannot rule out that such frequency-dependent deviations were perceptually relevant, especially in the present dynamic task. Thus, while the system was well-matched according to static and broadband acoustic metrics, the dynamic ADP task in virtual audio conditions may produce cues that differ slightly from real-world perception and affect distance judgments.

Relating our findings on the effect of auditory virtualization to results from previous studies is difficult, as no other study has systematically examined the impact of virtualization on dynamic ADP. Furthermore, our study examined dynamic ADP over exceptionally large distances, with the effects being strongly driven by the farthest loudspeakers (LS4 and LS5), which further limits direct comparisons with other studies usually examining closer distances. That said, Blau et al.~\cite{Blau2021} and Stärz et al.~\cite{starz_comparison_2025} found no significant bias in ADP when comparing real and virtual loudspeakers in 1DoF conditions. Conversely, Stodt et al.~\cite{stodt_comparing_2024} showed that detecting distance changes between loudspeakers was reduced in virtualized conditions using 0DoF binaural rendering, aligning somewhat with our findings. Lastly, similar to our findings at first glance, Lombera et al.~\cite{lombera_comparison_2025} found compressed distance responses or even strong underestimation for a subset of participants in 0DoF binaural rendering. However, compression and underestimation occurred only in incongruent room conditions, with significant differences in reverberation time between the room used to capture BRIRs and the reproduction room---a situation entirely unrelated to our study (see Fig.~\ref{fig:T20_DRR_LUFS}).

The results for conditions involving virtualized vision (RA/VV and VA/VV) offer additional insight into the role of visual context in dynamic ADP. With vision alone virtualized (RA/VV), accuracy decreased and effort increased relative to the real baseline, but signed errors showed only a non-significant trend toward a slight overestimation of distance. Thus, visual virtualization appears to increase uncertainty in mapping the auditory percept onto visible loudspeakers, while producing, at most, a weak directional shift. This trend is consistent with the visual distance-compression assumption: because egocentric visual distances are often underestimated in HMD-based VR~\cite{kelly_distance_2023, kurpinar_strengths_2025}, the visually rendered loudspeakers may have appeared slightly closer than their corresponding real acoustic positions. This could lead listeners to select a farther loudspeaker to match the perceived auditory distance, resulting in a positive signed error. However, the effect was small and not statistically significant, suggesting that the visually induced compression did not exceed the point at which the percept of a given loudspeaker would shift toward the next farther response option in a reliable way. Instead, the physically real dynamic auditory cues in RA/VV most likely dominated perceived distance, modulated by the virtual visual scene reducing response precision.

Importantly, and in regard to RQ4, the VA/VV condition shows that the effects of visual and auditory virtualization do not combine additively. Compared to the VA/RV condition, the fully virtual condition did not result in higher inaccuracies, a shift in bias, or higher effort. Therefore, the inaccuracies observed when only vision was virtualized in RA/VV did not simply carry over and accumulate in VA/VV. Rather, once audio was virtualized, performance in the dynamic ADP task appears to have been dominated by virtual auditory cues, suggesting that auditory virtualization determined the error patterns of the fully virtual condition. This may be due to a reduced weighting of the visual domain in the audio-visual ADP task because virtualization of visual information provided less reliable cues, as hypothesized above.

\subsection{Plausibility and Task-Specific Behavioral Validity}

A central motivation of this study was to examine whether a plausible XR environment reproduces behavioral outcomes comparable to those observed in the real world. The present results demonstrate that plausibility and behavioral validity are related but not equivalent. While the XR audio system was plausible for the majority of participants, especially in VR, systematic differences in dynamic ADP were observed between the real and virtualized conditions, with lower accuracy and a bias toward nearer loudspeakers with auditory virtualization.

However, the dynamic ADP results do not contradict the plausibility findings. Rather, the two experiments investigated slightly different perceptual processes. The plausibility experiment assessed whether listeners could differentiate between real and virtual sound sources using any available cues, while the ADP task required listeners to focus on specific perceptual attributes related to sound source distance and map those percepts onto a set of response alternatives while moving. Such a focused task may be sensitive to subtle cue differences that do not necessarily break the impression of a plausible auditory event in the more holistic plausibility evaluation.

For XR-based hearing science, the present study suggests that plausibility should be considered an important baseline criterion regarding the quality of an XR system, but not a sufficient validation measure with respect to task-specific behavioral validity of said system. Thus, plausibility assessment can confirm that the environment is free of obvious artifacts or inconsistencies and that listeners generally perceive the rendered events as near-realistic, but task-specific validation remains necessary when the goal is to study fundamental perceptual processes or generalize behavioral findings to real-world listening.

\subsection{Limitations and Future Work}

Several limitations should be considered when interpreting the results. The study was conducted in a single, large laboratory room with specific acoustic and geometric properties. Different rooms with different acoustics or more complex geometries may lead to different outcomes, as the relative importance of auditory and visual cues may change. Furthermore, the acoustic simulation employed an efficient shoebox approximation and non-individualized HRTFs. These choices were deliberate, as they enabled a lightweight, real-time-capable 6DoF XR system. However, simplifications of the room geometry, deviations in dynamic room acoustic cues, and interindividual differences in binaural cues may have contributed to discriminations between real and virtual sound sources in the plausibility task and reduced performance in the dynamic ADP task. Future studies should therefore examine whether similar behavioral results emerge with other room types, more geometrically detailed acoustic models, and individualized HRTFs.

The locomotion-based ADP task constrained participants to walk along a straight line in front of the loudspeakers. This allowed us to focus on dynamic level and DRR cues while minimizing directional cues. However, everyday listening involves richer trajectories and head movements, which may alter the use of dynamic cues and the relative influence of visual information. Furthermore, the dynamic ADP task employed categorical loudspeaker identification as an intuitive and straightforward response method. However, previous work has shown that the response method can affect distance estimates~\cite{roskopf_impact_2025}. Thus, future studies may examine dynamic ADP with richer movement patterns and compare different response methods within the same 6DoF XR system to further differentiate between perceptual effects and those introduced by movements or response methods.

Finally, participants mostly had self-reported normal hearing and were relatively experienced with audio and listening experiments. Future work could examine whether similar behavioral results occur for less experienced listeners. On the other hand, future studies should assess the applicability of such a state-of-the-art 6DoF XR system to individuals reporting functional hearing difficulties, as ultimately, an important and highly promising use case of this XR technology is the ecologically valid screening and treatment of functional hearing difficulties~\cite{Ramirez2024_TIH,ramirez_speech--noise_2024,ramirez_screening_2026}.

\section{Conclusion}

XR is a powerful technology for studying spatial hearing in controlled yet naturalistic conditions, but its usefulness for hearing science depends on whether virtualized sensory cues elicit perceptual behavior similar to that in the real world. This study investigated how auditory and visual virtualization affect plausibility and dynamic ADP in a real-time 6DoF XR environment. The implemented audio-visual system provided plausible binaural rendering in both AAR and VR, with higher plausibility observed in VR, likely because the virtual visual scene reduced participants' sensitivity to real-virtual differences. However, the locomotion-based ADP experiment showed that plausibility does not necessarily imply behavioral equivalence with the real world. Although overall performance remained high, all virtualized conditions led to reduced distance-estimation accuracy and increased effort relative to the fully real baseline. Visual-only virtualization reduced accuracy and increased effort without inducing a distance bias, whereas auditory virtualization produced a systematic bias toward nearer perceived source distances. Importantly, performance in the fully virtual condition was primarily defined by auditory virtualization rather than by an accumulation of auditory and visual virtualization effects, suggesting that effects of auditory and visual virtualization do not add up. 

Overall, these results demonstrate that current 6DoF XR systems can provide environments that are both plausible and behaviorally highly useful for hearing science. However, they also reveal that plausibility alone may not be sufficient to establish real-world perceptual equivalence. For XR-based hearing science, this implies that immersive systems should be validated not only for overall plausibility but also for the specific perceptual and behavioral processes they are intended to study, diagnose, or train.

\backmatter

\section*{Declarations}

\bmhead{Supplementary information}
The supplementary material for this article~\cite{supp_material} will be made publicly available upon publication of the peer-reviewed article. Please see the ``Data availability'' declaration for details.
\bmhead{Acknowledgments}
The authors would like to thank all the voluntary participants for their support.
\bmhead{Funding}
This work was supported in part by the German Research Foundation (Project number: 444873001).
\bmhead{Conflict of interest}
The authors declare no conflict of interest.
\bmhead{Ethics approval and consent to participate}
All participants provided written informed consent for their voluntary participation in the study and for the subsequent publication of the results. All procedures were approved by the Research Ethics Board of Leipzig University (2020.07.21\_eb\_52) in accordance with the Declaration of Helsinki.
\bmhead{Consent for publication}
All authors have read and approved the final version of the manuscript for publication.
\bmhead{Data availability}
The supplementary material~\cite{supp_material}, which contains additional photos, screenshots, figures, video material, experimental data, evaluation code, and Jamovi files, was submitted together with the manuscript for peer review. Upon publication of the peer-reviewed article, the supplementary material will be permanently archived in a public Zenodo repository and linked from the published version of this article.
\bmhead{Author contribution}
J.M.A.: Conceptualization, Methodology, Software, Validation, Formal Analysis, Investigation, Data Curation, Writing - Original Draft, Writing - Review \& Editing, Visualization, Supervision, Project Administration. C.S.: Methodology, Software, Validation, Investigation, Writing - Review \& Editing. M.R.: Conceptualization, Methodology, Writing - Review \& Editing. S.W.: Conceptualization, Resources, Writing - Review \& Editing, Supervision, Project administration, Funding acquisition. F.B.: Conceptualization, Methodology, Formal Analysis, Writing - Review \& Editing, Supervision, Project administration, Funding acquisition.

\bibliography{Lib_preprint}

\end{document}